\documentclass[final,3p,11pt,authoryear]{elsarticle}
\pdfoutput=1 

\journal{Journal of Fluids and Structures}

\usepackage{graphicx,color,amsmath,array,amssymb,amsthm,parskip}
\usepackage{multirow}

\begin{document}

\begin{frontmatter}



\title{Phase dynamics of effective drag and lift components in vortex-induced vibration at low mass--damping}


\author[aff1]{E. Konstantinidis}
\address[aff1]{Department of Mechanical Engineering, University of Western Macedonia, Bakola and Sialvera, Kozani 50132, Greece}

\author[aff2]{J. Zhao}
\address[aff2]{Department of Mechanical and Aerospace Engineering, Monash University, Melbourne, VIC 3800, Australia}

\author[aff4]{J. Leontini}
\address[aff4]{Department of Mechanical Engineering and Product Design Engineering, Swinburne University of Technology, John St Hawthorn, 3162, Australia}

\author[aff3]{D. Lo Jacono}
\address[aff3]{Institut de Mécanique des Fluides de Toulouse (IMFT) CNRS, UPS, Université de Toulouse, Allée Camille Soula, F-31400 Toulouse, France}

\author[aff2]{J. Sheridan}

\begin{abstract} 
In this work, we investigate the dynamics of vortex-induced vibration of an elastically mounted cylinder with very low values of mass and damping. We use two methods to investigate this canonical problem: first we calculate the instantaneous phase between the cylinder motion and the fluid forcing; second we decompose the total hydrodynamic force into drag and lift components that act along and normal to, respectively, the instantaneous effective angle of attack. We focus on the phase dynamics in the large-amplitude-response range, consisting of the initial, upper and lower ``branches'' of response.  The instantaneous phase between the transverse force and displacement shows repeated phase slips separating periods of constant, or continuous-drifting, phase in the second half of the upper branch.  The phase between the lift component and displacement shows strong phase locking throughout the large-amplitude range -- the average phase varies linearly with the primary frequency -- however the modulation of this phase is largest in the second half of the upper branch. These observations suggest that the large-amplitude-response dynamics is driven by two distinct limit cycles -- one that is stable over a very small range of reduced velocity at the beginning of the upper branch, and another that consists of the lower branch. The chaotic oscillation between them -- the majority of the upper branch -- occurs when neither limit cycle is stable. The transition between the upper and lower branches is marked by intermittent switching with epochs of time where different states exist at a constant reduced velocity. These different states are clearly apparent in the phase between the lift and displacement, illustrating the utility of the force decomposition employed.  The decomposed force measurements also show that the drag component acts as a damping factor whereas the lift component provides the necessary fluid excitation for free vibration to be sustained.
\end{abstract}


\end{frontmatter}


\section{Introduction}\label{sec:intro}
Vortex-induced vibration (VIV) has been the subject of extensive research over the past six decades because of  its importance in engineering applications, such as riser pipes  transporting oil from the sea bottom, supporting cables and pylons of offshore platforms, on one hand and because of the complexity of the fluid-mechanical phenomena on the other hand. There are several review papers on the subject  including \cite{Parkinson1971}, \cite{King1977}, \cite{Griffin81b}, \cite{Bearman1984}, \cite{Sarpkaya04},\cite{Williamson04}, \cite{Gabbai2005} and \cite{Bearman2011}.  Much of the fundamental research has dealt with rigid circular cylinders  elastically mounted so as to have one degree of freedom to oscillate transversely to a uniform free stream. This can be considered as the simplest configuration to study VIV and the building block to understand phenomena in more complex configurations \citep{Williamson04}. Yet, semi-empirical codes and guidelines used in the industry also rely on databases of the hydrodynamic forces on rigid cylinders undergoing single degree-of-freedom transverse oscillations. 
The equation of motion for a freely-vibrating hydro-elastic cylinder can be written as
\begin{equation}\label{eq:motion1}
	m\left(\ddot{y} + 2\pi f_{n}\zeta\dot{y} + 4\pi^2f_n^2 y\right) =  F_y,
\end{equation}
where $y$ is the displacement of the cylinder and each overdot represents a derivative with respect to time, $m$ is the mass of the oscillating structure, $f_n$ and $\zeta$ respectively are the natural frequency and the damping ratio of the structure both measured in vacuum, and $F_y$ is the time-varying hydrodynamic force acting in the direction of motion. In the above form, the left-hand side of the equation of motion is a second-order differential equation that comprises parameters exclusively associated with the solid structure. Equation (\ref{eq:motion1}) shows that the structural dynamics is linear and any non-linearity is introduced by the hydrodynamic forcing term on the right-hand side. As a consequence, the modelling of the hydrodynamic force $F_y$ has significant ramifications for understanding VIV, which is the main theme of the present work.

\subsection{Experimental characterisation of the response}
In experimental studies, the natural frequency  and the damping ratio  are typically determined from free-decay tests in still fluid \citep{Blevins01}.  Measured values  from free-decay oscillations in still fluid differ from the true values, which correspond to the solid structure  in vacuum \citep{Sarpkaya04}.  \cite{Blevins09b} has provided a hydrodynamic model  for oscillations of a cylinder in still fluid that can be used to estimate the true values from free-decay oscillations in still fluid.  
For tests in air, which  has density  much lower than the average density of solid structures, measured values from free-decay oscillations can be regarded to a good approximation as the true values. 
For tests in water or fluids with a density comparable to the average density of the structure, the surrounding fluid alters the measured natural frequency and damping ratio. In numerical simulations, the true values in vacuum are almost exclusively set as input parameters \citep[see, e.g., ][]{Lucor2005,Leontini2006}. The dimensional parameters governing the VIV response are typically combined into four dimensionless groups: the mass ratio $m^*$ based on the dry mass of the structure, the damping ratio $\zeta$ estimated directly from free-decay measurements  in still air (nominally the same as in vacuum), the reduced velocity $U^*$ based on the natural frequency of the structure in still fluid, and the Reynolds number $Re$ based on the cylinder diameter and free-stream speed. 

For hydro-elastic cylinders with very low values of mass and damping so that the product $m^*\zeta$ is of the order of 0.01 or less, \citet{Khalak96,Khalak99} identified four distinct branches of response with changing $U^*$: the initial excitation region, the upper branch of very high amplitude, the lower branch of moderate amplitude, and the desynchronisation region. Transitions between response branches involve jumps in the cylinder oscillation amplitude and are accompanied by changes in the mode of vortex shedding \citep{Govardhan00}. For hydro-elastic systems with high values of combined mass–damping, the upper branch of very high response amplitude does not appear as found in early experiments by  \citet{Feng1968} and more recently by \citet{Klamo06}. A very recent study showed that the upper branch does not disappear  at high levels of damping but  jumps in response amplitude become smeared out instead;  the  upper branch still exists as can be inferred from the variation of the response frequency \citep{Soti18}. Both upper and lower branches have been associated to synchronisation, i.e.\ the oscillation and the fluid forcing synchronise at a common frequency. However, the common frequency increases in the upper branch but remains fairly constant in the lower branch, which indicates that the dynamics is different in these two branches.

\subsection{Lock-in and synchronisation}
Lock-in is the phenomenon where the frequency of vortex shedding becomes self-tuned to approximately the natural frequency  of the hydro-elastic cylinder for a range of free-stream velocities. It should be noted that we define `lock-in' in a manner that is not tantamount to `synchronisation' \citep[see also][]{Konstantinidis2014}. Since the classical work of \cite{Feng1968} the phenomenon was considered to be the result of the structural motion dominating the vortex formation process thereby controlling the frequency at which vortices are shed in the wake. \cite{Govardhan00} showed that classical lock-in is observed in the lower branch. They showed that the lock-in frequency depends on the mass ratio according to the following formula,
\begin{equation}\label{eq:GWformula}
	f^*_\text{lower} = \sqrt{\frac{m^*+1}{m^*-0.54}},
\end{equation}
obtained from an empirical  fit to experimental data with different mass ratios for very low values of $m^*\zeta$. The recent measurements of \citet{Soti18} showed that the frequency  of cylinder response is constant in the lower branch also at high levels of damping. On the other hand, \citet{Gharib1999} did not observe lock-in behaviour in his experiments for mass ratios below 10. Experimental tests by \citet{Blevins09a} showed that lock-in tendency weakens with increasing either mass or damping of the hydro-elastic cylinder. A lock-in region of fairly constant frequency of cylinder response is also not evident in the experiments of \citet{Lee2011}, who conducted extensive tests with variable levels of damping and stiffness, which were adjusted by a virtual damper-spring apparatus.

We note here the important difference between the notions of lock-in and synchronisation as we have employed it in this study. Here, we use the definition of synchronisation of weakly-coupled oscillators given by \citet{Pikovsky2001} - that two oscillators are synchronised if their interaction causes their mean frequencies to match. This means that two oscillators can be synchronised while each oscillates chaotically - in such a case, it would be expected that the phase lag between the two oscillators would only vary by a small amount over time. We assume that the cylinder-wake system can be treated in this way, as two separate oscillators whose dynamics are weakly coupled and therefore susceptible to synchronisation. In this framework, lock-in is clearly an example of synchronisation, but the concept applies to a wider set of responses, potentially including non-harmonic and chaotic oscillations.

\subsection{Linearised modelling - the assumption of harmonic oscillations}
On the theoretical side, much of our understanding of VIV relies on the approximation of pure harmonic oscillation \citep{Bearman1984,Bearman2011}. That is, assuming that the displacement of the cylinder can be described as a sinusoidal function of time $t$,
\begin{equation}\label{eq:sinusoid1}
	y(t)=A\sin{(2\pi ft)},
\end{equation}
where $A$ and $f$ respectively are the amplitude and the frequency of oscillation. By taking the first and second order derivatives of (\ref{eq:sinusoid1}) and  inserting into (\ref{eq:motion1}), it follows directly that the hydrodynamic force $F_y$ is also a pure harmonic function of time, which is typically expressed in terms of the force coefficient $C_y$ and the phase lag $\phi$ as  
 \begin{equation}\label{eq:totalforce}
	F_y(t)  =  \frac{1}{2}\rho U_\infty^2 DL C_y\sin{\left(2\pi ft+\phi\right)},
\end{equation}
where $U_\infty$ is the free-stream velocity, $D$ and $L$ are the cylinder diameter and immersed length. In the following, it is assumed that the hydrodynamic force and the oscillation are homogeneous along the spanwise direction so that it is permissible to consider a unit length of the cylinder. 

By replacing the assumed harmonic motion (\ref{eq:sinusoid1})  and hydrodynamic force (\ref{eq:totalforce}) in the equation of motion (\ref{eq:motion1}) and equating the factors of the sine and cosine terms on both sides, a set of two algebraic equations is obtained. 
\cite{Khalak99}, among other investigators,  solved the set of equations for the response amplitude and frequency and obtained the following relationships:
\begin{equation}\label{eq:Astar}
	A^*  = \frac{C_y\sin\phi}{4\pi^3(m^*+C_A)\zeta_a}\left(\frac{U^*}{f^*}\right)^2f^*,
\end{equation}
\begin{equation}\label{eq:fstar}
	f^* = \sqrt{\frac{m^*+C_A}{m^*+C_{EA}}},
\end{equation}
where $C_{EA}$ is an `effective added mass coefficient' that depends on the  component of the hydrodynamic force in-phase with the cylinder displacement:
\begin{equation}\label{eq:CEA}
	C_{EA}  = \frac{C_y\cos\phi}{2\pi^3A^*}\left(\frac{U^*}{f^*}\right)^2.
\end{equation}
In the above equations that are written in dimensionless form, $A^*=A/D$, $f^*=f/f_{n\text{,fluid}}$,  $m^*=4m/\pi\rho D^2L$, $\zeta_a=c/2\sqrt{k(m+m_A)}$ and $U^*=U_\infty/f_{n\text{,fluid}}D$, where $f_{n\text{,fluid}}$ is the natural frequency measured in still fluid,  $c$ is the structural damping, and $m_A=C_A\pi\rho D^2L$ is the added mass. The added mass coefficient $C_A$ for small amplitude oscillations in otherwise still fluid takes a value of 1.0 \citep{Williamson04}. The above set of equations shows that the normalized amplitude and frequency of response depends on the mass ratio $m^*$, the damping ratio $\zeta_a$, and the reduced velocity $U^*$. These dependencies have been relatively well established from experimental observations where $U^*$ is typically varied over the range of interest.  

The force coefficients  in-phase with cylinder displacement $C_y\cos\phi$ and velocity $C_y\sin\phi$, which appear in the solution of the linearised problem, depend on the motion of the cylinder in a highly complex manner. A line of thought is to derive the force coefficients from measurements using controlled harmonic oscillations \citep[see, e.g.,][]{Gopalkrishnan93,Morse09a}. The force coefficients are typically taken as Fourier averages over many cycles of oscillation of the transverse component of the unsteady force acting on the cylinder. These measurements can be used in conjunction with the assumption of harmonic oscillation  to predict the free response of an elastically mounted cylinder. \cite{Morse09b} found that matching the Reynolds number between controlled and free vibrations is important for successfully predicting the peak oscillation amplitude in the upper branch.  \cite{Blevins09b} adopted the converse approach: in this study the force coefficients were computed from the harmonic model equations for free oscillations of a cylinder transverse to a free stream; both steady and transient conditions were employed to cover a parameter space of normalized amplitude and frequency of interest. Then, the constructed force database was used to to predict the response for systems with different levels of structural damping.  

The component of the fluid force in-phase with the cylinder velocity $C_y\sin\phi$, often referred as  the excitation coefficient, can sustain self-excited oscillations if it is positive, a requirement that stems directly from (\ref{eq:Astar}). In the special case of zero structural damping $(\zeta=0)$, which is often employed in numerical simulations, $C_y\sin\phi$ has to be zero. The excitation coefficient $C_y\sin\phi$ also represents the normalized energy transfer from the fluid to the cylinder motion over an average cycle \citep{Morse09b}. Several investigators employed controlled harmonic motion to identify the regions of positive energy transfer where free harmonic vibration is possible \citep{Tanida1973,Hover1998,Carberry2005}. As shown by \cite{Morse09b,Morse10}, a very large number of tests is necessary in order to precisely map the regions of positive energy transfer in the parameter space of normalized amplitude and wavelength $\{A^*:U^*/f^*\}$ and replicate the dynamics of free vibration in different response branches.  The latter investigators further noted that an additional requirement for the success of the prediction of free vibration is the stability of the harmonic solutions. 

\subsection{Deviations from harmonic oscillations}
There is some debate whether the assumption of harmonic motion is  a good approximation of VIV under all circumstances. \citet{Sarpkaya04} discussed possible limitations of this linearised approach when the oscillations have amplitude and/or frequency modulations. \citet{Marzouk2011} carried out numerical simulations at low Reynolds numbers and found that  a constructed forced vibration that does not contain the third-superharmonic component of the main oscillation frequency failed to reproduce some details of the fluid force although the magnitude of this component is about 400-fold less than the magnitude of the fundamental component.  This was significant at low values of damping and became less significant at high damping values for a hydro-elastic cylinder with a mass ratio of unity. \cite{Konstantinidis2017} noted, by examining data from both controlled and free vibrations  at a Reynolds number of 100 from numerical simulations of other investigators, that VIV has to deviate from pure harmonic since some operating points of free oscillation for a system without damping do not fall atop the contour of zero energy transfer  in the parameter space of normalised amplitude and frequency.  \citet{Zhao2014a} accurately measured the free vibration of a hydro-elastic  cylinder and used the recorded displacement signal to drive a cylinder in controlled motion replicating the free vibration; they found that the controlled motion did not replicate the same flow patterns and the same phasing of the hydrodynamic force observed in free vibration at a point near peak oscillation amplitude in the upper branch, suggesting chaotic, and therefore clearly non-harmonic, oscillations.

\subsection{The link between phase and the mode of vortex shedding}

\subsubsection{Understanding phase from a harmonic oscillation perspective}
The accurate measurement of the phase lag $\phi$  between fluid force $F_y(t)$ and  displacement $y(t)$  is critical for analysing and  interpreting fluid forcing data. This is so because the phase of the fluid force is generally considered to be influenced by the character or mode of vortex shedding \citep{Gabbai2005,Bearman09}. For controlled harmonic oscillation of a cylinder transversely to a free stream, \citet{Morse09a}  have provided high-resolution contours of $\phi$ in the parameter space of normalised amplitude and wavelength $\{A^*:U^*/f^*\}$. They showed that contours are not continuous across the entire parameter space but distinct boundaries appear where changes in the regime of vortex shedding were identified by flow visualisation.  However, the relationship between fluid force phase and vortex shedding mode remains rather unclear due to the lack of an analytical model that can explain physically and quantitatively the variations in $\phi$ within each regime.  In free vibration, variations in the phase of the fluid force are even more difficult to interpret due to restrictions posed by the equation of the cylinder free motion. Assuming that free vibration is pure harmonic, \citet{Sarpkaya04} derived the following relationship : 
\begin{equation}\label{eq:phase}
	\tan\phi = \frac{2\zeta ff_n}{f_n^2-f^2}. 
\end{equation}
As a consequence, $\phi $ must change sign at $f=f_n$, which brings in an extra dependency on the mechanical properties of  hydro-elastic cylinder in addition to the dependency on the amplitude and frequency of cylinder motion.  Equation (\ref{eq:phase}) shows that $\phi$ is only a function of the damping ratio $\zeta$ and the frequency ratio $f/f_n$ irrespectively of the mass ratio $m^*$ and the reduced velocity $U^*$. At very low levels of damping $\zeta$ of the order of $10^{-3}$ in particular, $\phi$ is constrained to values close to $0^\circ$ or $180^\circ$ and a phase shift by approximately $180^\circ$ occurs at $f=f_n$. With such a limited range of  permissible $\phi$ values, it becomes hard  to unambiguously discern changes in the mode of vortex shedding. 

\subsubsection{Understanding phase when harmonic oscillation is not present}
The phase lag between force and displacement can also change during a test at some fixed reduced velocity.  By employing the Hilbert transform to compute the instantaneous phase during free vibration, \citet{Khalak97a}  showed that for operating points in the transition region from the upper branch to the lower  branch  intermittent switching of the instantaneous phase between $0^\circ$ and $180^\circ$ occurs. Their finding highlights that the use of the mean phase may be less representative in the transition region,  at least.

Of course, it should be noted that an instantaneous switch in phase does not imply an instantaneous switch in vortex wake organisation or mode. A change in mode  may require several cycles of oscillation to complete. Phase switching appears to be more sudden than is physically possible for vortex shedding to switch from one side to the other side of an oscillating cylinder as a $180^\circ$ jump in phase implies. Therefore, using phase only as an indicator of vortex shedding mode can only be suggestive.

With this caveat in mind, \citet{Zhao2014a} employed the Hilbert transform to show the phase dynamics, and use the statistics of the phase fluctuation to understand the vortex dynamics in the wake of a freely-vibrating cylinder. They  confirmed that the upper$\leftrightarrow$lower branch  transition is associated with intermittent switching of the instantaneous phase.  Although it was clear that different vortex patterns or wake modes occurred during a test, the relationship to intermittent switching in the phase of the unsteady force with respect to the displacement may require further investigation.

Furthermore,  \citet{Hover1998} observed  a significant drop in the correlation coefficient of  the measured forces at the two ends of a freely-vibrating cylinder in part of the upper branch. This could be indicate inhomogeneity of the vortex shedding process along the cylinder span.  In approximately the same region, \citet{Zhao2014a} found evidence of chaos caused by mode competition between at least two distinct  modes of vortex shedding. One of the objectives of the present  study is to examine the origin of phenomenologically chaotic dynamics observed in the upper branch. 

 \subsection{An outline of the approach of this study}
\begin{figure}
\begin{center}
\includegraphics[width=0.7\textwidth]{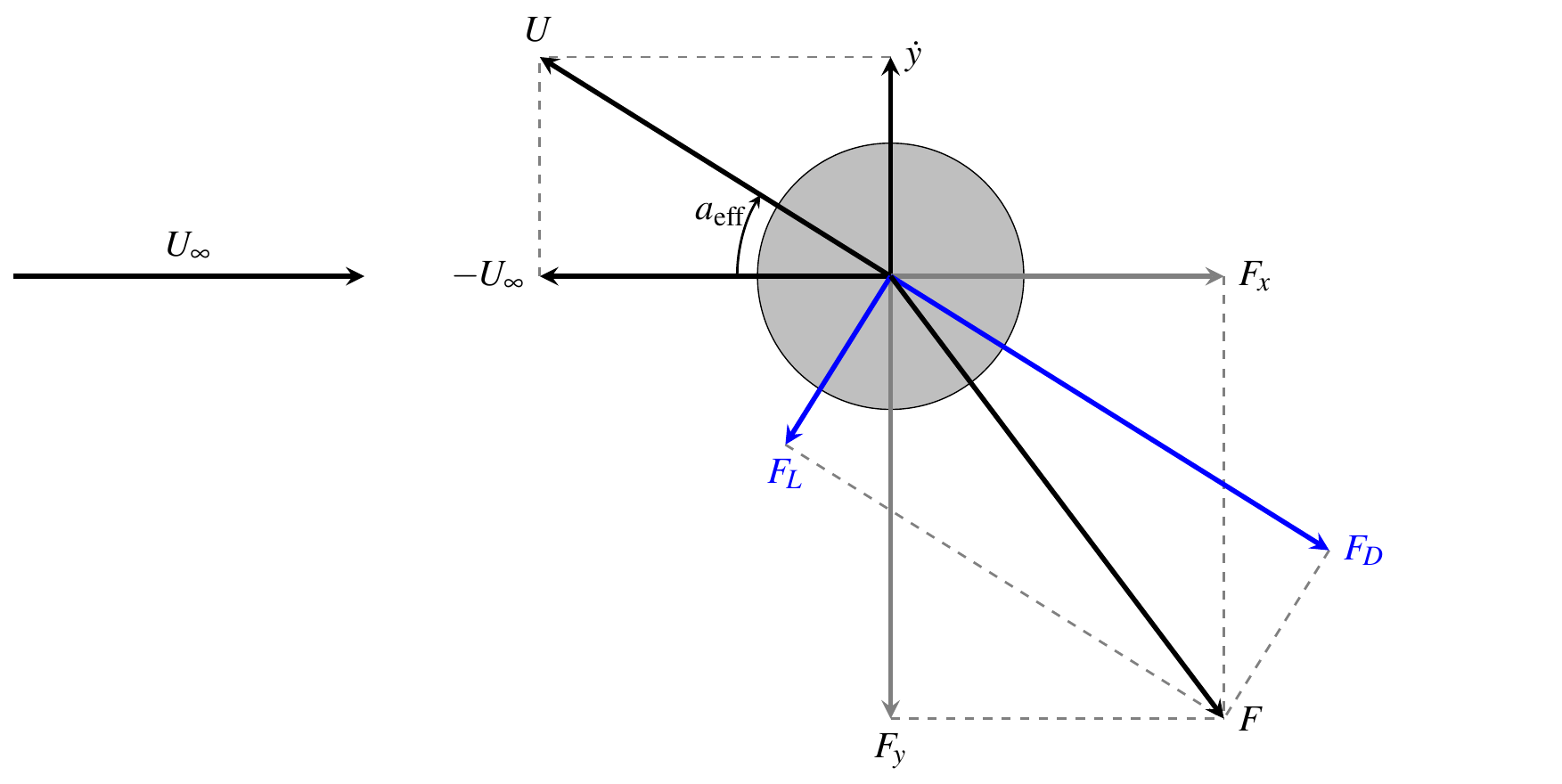}
\end{center}
\caption{Diagram of the main vector quantities employed in the present study in the relative frame of reference moving with velocity $U$ as the cylinder moves through still fluid.} \label{fig:vector_diagram}
\end{figure}

In this study, we approach VIV using an alternative perspective. As a cylinder oscillates transversely to a free stream, the effective angle of attack changes. The relative velocity between the free-stream velocity $U_\infty$ and the cylinder velocity $\dot{y}$ is $\boldsymbol{U}=U_\infty\boldsymbol{i}-\dot{y}\boldsymbol{j}$, where $\boldsymbol{i}$ and $\boldsymbol{j}$ respectively are the unit vectors in the streamwise $x$ and transverse $y$ directions. Here, we consider the kinematically analogous case where the cylinder is towed through still fluid with its axis maintained perpendicular to the direction of motion \citep{Konstantinidis2013}. The cylinder moves forwards at constant speed $U_\infty$ while its speed in the transverse direction changes. We define the `effective' drag $F_D$ and lift $F_L$ as the components of the total hydrodynamic force $F$ acting along the instantaneous angle of attack $a_\text{eff}$ and normal to the drag direction, respectively,  as shown in figure~\ref{fig:vector_diagram}. In the following, drag and lift are per their definition in figure~\ref{fig:vector_diagram}. The components of the force in the horizontal direction $F_x$ and the vertical direction $F_y$ are those customarily measured in the fixed frame of reference of the laboratory. 

The selection of a relative frame of reference has some important implications. As the cylinder is towed in a prescribed path through still fluid, it does work on the fluid at the rate $\boldsymbol{F}\cdot\boldsymbol{U}$  where $\boldsymbol{F}$ and  $\boldsymbol{U}$ respectively are the instantaneous vectors of the hydrodynamic force and the cylinder velocity.  It is reasonable to assume that the fluid's drag force resists the motion of the cylinder, i.e.\ it is a reactive force (note that we present results in section \ref{sec:indirect} that shows this is indeed the case). Thus, only the force aligned with the instantaneous velocity of the cylinder in the relative reference frame can transfer energy from the fluid to the cylinder. This must also be true in the fixed reference frame since the two cases are kinematically equivalent. In the case of an elastically-mounted cylinder undergoing free vibration transversely to a uniform free stream, the energy required to excite and sustain the oscillations can only come from the lift component, which therefore may be directly linked to the vortex dynamics in the wake. We use this approach in an attempt to gain insight into the dynamics of VIV and answer outstanding questions posed in the foregoing paragraphs.

\section{Experimental set-up}
\begin{figure}
\begin{center}
\includegraphics[width=0.75\textwidth]{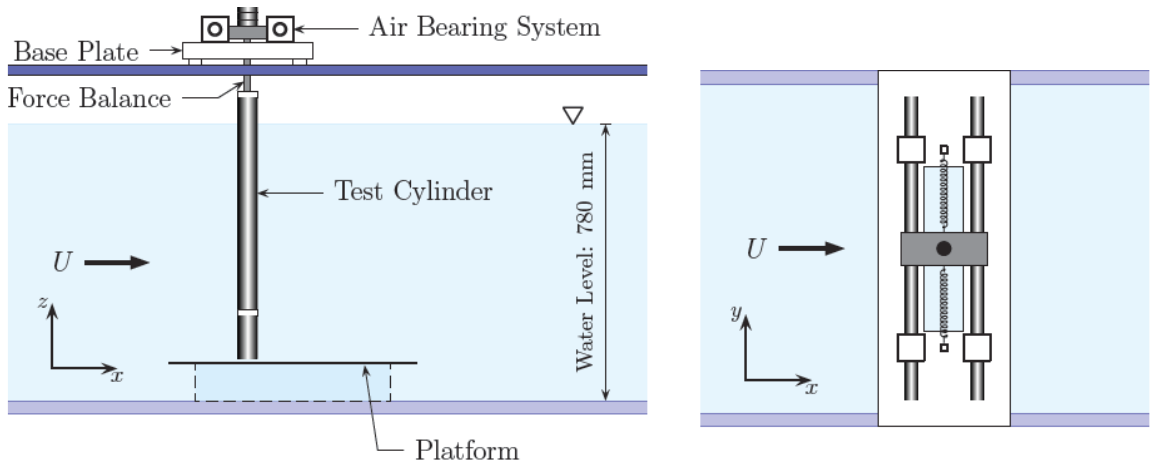}
\end{center}
\caption{Schematic of the experimental facility. }\label{fig:schematic}
\end{figure}

The experiments were conducted in a free-surface recirculating water channel  of the FLAIR group at Monash University. Figure~\ref{fig:schematic} shows a schematic of the experimental configuration. The water channel has a test section of 0.6 m (width) $\times$ 0.8m (height) $\times$ 4m (length) and the background turbulence level is below 1\%. A rigid cylinder made of carbon fibre tubing was elastically mounted from above the free surface on low-friction air bearings. The cylinder had an outer diameter of $D = 25$ mm and an immersed length of $L = 620$ mm, giving an aspect ratio of $L/D = 24.8$. A raised platform was placed at the bottom of the water channel with a gap of approximately 1 mm to the free end of the cylinder  to promote two dimensionality of the flow along the span.  The  ratio of the mass of the oscillating structure to the mass of the fluid displaced by the cylinder was estimated to be $m^*=3.00$. The natural frequency of the mechanical system was measured from free-decay tests, which resulted in values of $f_{n,\text{air}} = 0.835$ Hz in still air and $f_{n,\text{water}} = 0.717$ Hz in still water. The ratio of the mechanical damping to the critical damping of the mechanical oscillator was estimated from the free-decay tests in air to be $3.5\times10^{-3}$. In this study, it is assumed that the mechanical properties measured in still air approximately correspond to the properties of the system in vacuum. The free-stream velocity was increased  from 43.1 to 260.4 mm/s  at 96 different values leaving sufficient time between measurements for  conditions to settle. The adjustment of the free-stream velocity resulted in variations in the reduced velocity, defined as $U^*=U_\infty/f_{n,\text{water}}D$, between 2.4 and 14.5 corresponding to Reynolds numbers in the range from 1250 to 7550. It should be noted that in the presentation of the results we employ the reduced velocity based on the natural frequency of the structure in still water for consistency with previous studies. 

The displacement of the oscillating cylinder was monitored using a non-contact magnetostricitive linear variable differential transformer having an accuracy within 0.01\% of the 250 mm full-scale range, giving a displacement measurement precision of approximately $0.001D$. 
The streamwise and transverse components of the fluid force acting on the oscillating cylinder were simultaneously measured using a two-component force balance having a resolution of 0.005~N. The measurement of the force was based on strain gauges configured in a full Wheatstone bridge circuit. For the transverse component, the inertial force due to the cylinder’s acceleration was subtracted to recover the instantaneous fluid force. 
The force balance is designed to have a  natural frequency of the first mode shape at  1268.5 Hz, which is far greater than that of the hydro-elastic cylinder and the frequency of vortex shedding in the experiments ($<$ 2 Hz). 
Time series of the displacement, $y(t)$, streamwise, $F_x(t)$, and transverse, $F_y(t)$, components were collected over 300 s at a sampling rate of 100 Hz yielding $3\times10^4$ samples per channel. 
 
Details about the experimental set-up and measurement techniques can also be found in \cite{Nemes2012} and \cite{Zhao2014a,Zhao2014b}. A detailed comparison of response characteristics of rigid cylinders undergoing VIV obtained with the experimental facility used in this study, but with a different set of measurement instruments,  against previous experiments from the published literature has been given in \cite{Zhao2014b} and \citet{Soti18}. Overall, a good match has been found in terms of both amplitude and frequency variation with reduced velocity, which provides validation for the experimental set-up employed in the present study. 

\section{Data processing}\label{sec:processing}
The components of the hydrodynamic force along and normal to the free-stream direction, $F_x$ and $F_y$ respectively, were measured with the force balance. The drag, $F_D$,  and the lift, $F_L$,  components of the total force were obtained by a transformation from the laboratory (fixed) frame of reference to one attached to the cylinder as it would move with the same relative velocity through still fluid, using the following formulas
\begin{align}
	F_D = F_x\cos{a_\text{eff}} - F_y\sin{a_\text{eff}},\\
	F_L = F_x\sin{a_\text{eff}} + F_y\cos{a_\text{eff}},
\end{align}
where $a_\text{eff}=\tan^{-1}{\left( \dot{y}/U_\infty\right)}$  defines the instantaneous effective angle of the relative velocity vector (see figure~\ref{fig:vector_diagram}). The instantaneous velocity of the cylinder was computed by numerical differentiation of the time series of the displacement, $\dot{y}(t)=dy/dt$. The signal from the displacement sensor was first low-pass filtered to avoid propagation of errors due to measurement noise in the computed velocities. 

The time series of directly and indirectly measured  quantities (displacement and forces) were further processed to obtain their instantaneous attributes using the Hilbert transform as follows.  Taking an arbitrary signal $s(t)$  and its Hilbert transform $\hat{s}(t)$, the signal can be represented analytically as \citep{Cohen1995}
\begin{equation}
	s_A(t) = s(t) + \mathrm{i}\,\hat{s}(t) = A_s(t)e^{\mathrm{i}\phi_s(t)},
\end{equation}
where 
$$
		 A_s(t) = \sqrt{s^2(t) + \hat{s}^2(t)} \quad\text{and}\quad 
		\phi_s(t) = \arctan\left(\frac{\hat{s}(t)}{s(t)} \right). \eqno{(\theequation{\mathit{a},\mathit{b}})}\label{eq:hilbert}
$$
In principle,  $A_s(t)$ is the instantaneous amplitude and $\phi_s(t)$ is the instantaneous phase of the signal. Then, the instantaneous `monocomponent' frequency of the signal can be obtained as
\begin{equation}
	f_s(t) = \frac{1}{2\pi}\frac{\text{d}\phi_s}{\text{d}t}. 
\end{equation}
Particular attention was exerted to the application of the Hilbert transform on each signal, which was padded with portions of the signal itself to decrease end effects. In addition, it is important to remove any mean component of the original signal before applying the Hilbert transform, otherwise $\hat{s}(t)$ becomes spurious; this particularly concerns the drag component of the force. 

\begin{figure}
\begin{center}
\includegraphics[width=0.7\textwidth]{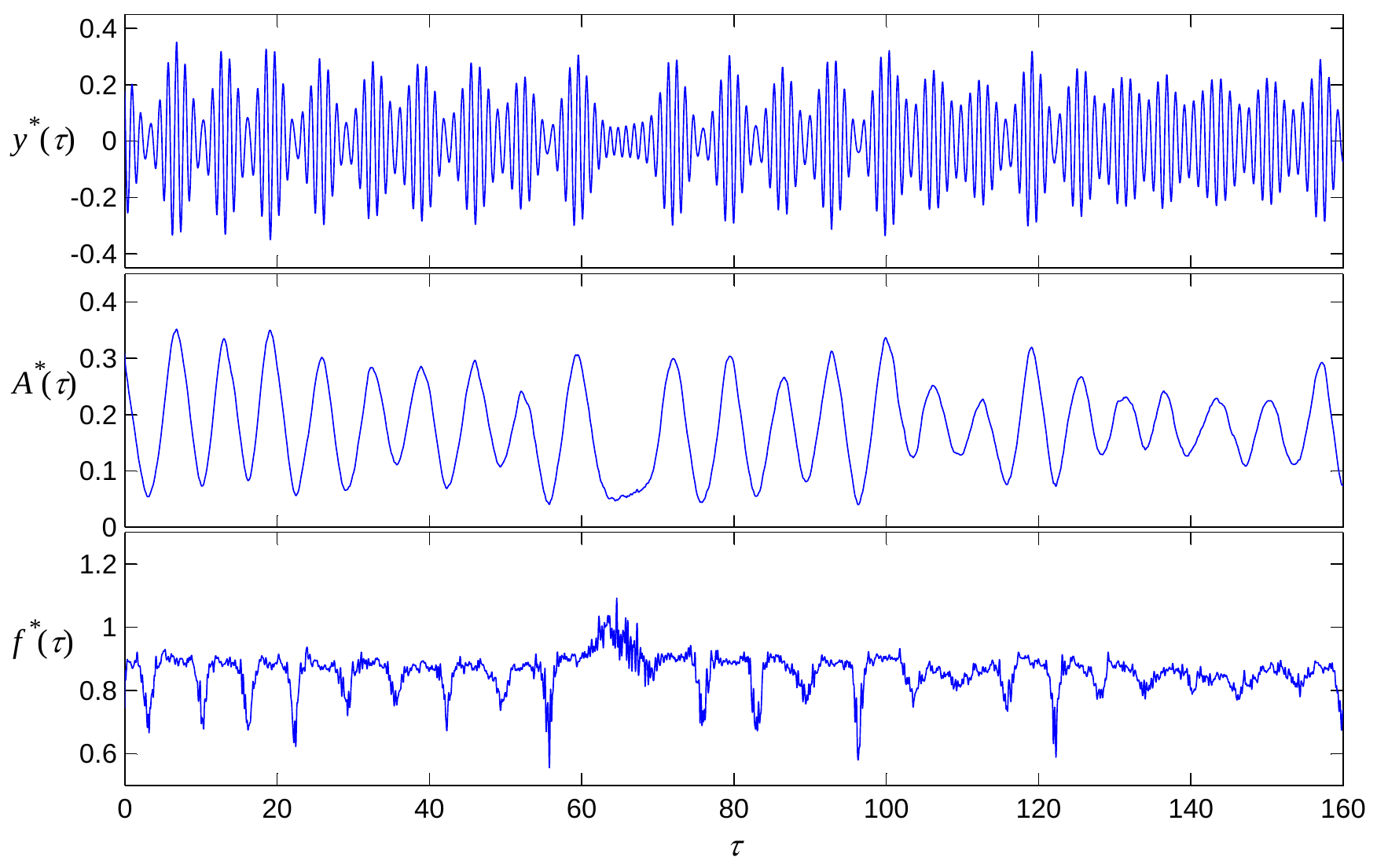}
\end{center}
\caption{Time series of the displacement, $y^*(\tau)$, its instantaneous amplitude, $A^*(\tau)$, and its instantaneous frequency, $f^*(\tau)$, obtained from the Hilbert transform at a reduced velocity of $U^*=4.2$. Asterisks denote normalisation using the cylinder diameter and the natural frequency in still water.}\label{fig:hilbert}
\end{figure}

As an example of the data-processing method, figure~\ref{fig:hilbert} shows how the Hilbert transform operates on the displacement signal, $y^*(\tau)=y/D$, to yield the instantaneous amplitude of cylinder vibration, $A^*(\tau)=A/D$, and the instantaneous frequency of vibration, $f^*(\tau)=f/f_{n,\text{water}}$, where $\tau=tf_{n,\text{water}}$ is the normalised time.  We have selected a special case at $U^*=4.2$, where the displacement signal exhibits strong  modulations,  which are reflected in the variation of the instantaneous amplitude and frequency.  In this case, the displacement $y^*(\tau)$ exhibits oscillations at two close frequencies, which correspond to approximately the frequency of vortex shedding from a fixed cylinder and the natural frequency of the cylinder in still water. The competition between these two frequencies leads to a quasi-periodic response, as also observed by \citet{Khalak99}.  As a consequence,  the instantaneous amplitude $A^*(\tau)$ of the cylinder response displays fluctuations at the frequency of amplitude modulation, which is equal to the difference between the competing frequencies. 
In the presentation of results ( e.g., see figure~\ref{fig:response}), we primarily report mean values of the vibration amplitude $A^*$, which have been averaged over the duration of each experimental run. Although mean values are not fully representative in particular cases where the response is quasi-periodic, we clearly point out such cases in the discussion of the results.
Similarly, we report mean values for other quantities,  which have been also obtained by averaging the instantaneous amplitudes of the corresponding quantities obtained from the Hilbert transform. More specifically, $C_y$ denotes the mean amplitude of the unsteady force coefficient in the transverse direction while $C_D$ and $C_L$ denote the mean amplitudes of the unsteady drag and lift coefficients along and normal to the effective angle of attack. 

The present analysis focuses on the phase dynamics, which refers to the  phase lag of the measured forces with respect to a reference signal, which is either the cylinder displacement or the relative velocity. Phase lags were calculated by taking the difference of the instantaneous phases of the signal and of the reference signal, which were both obtained from the Hilbert transform. For example, the instantaneous phase lag, $\tilde{\phi}(t)$, between the transverse force, $F_y(t)$, and cylinder displacement, $y(t)$, is computed as $\tilde{\phi}(t)=\tilde{\phi}_{F_y}(t)-\tilde{\phi}_{y}(t)$, where tildes denote the instantaneous phases. For the sake of simplicity, tildes will be dropped hereafter. Generally, we wrap the instantaneous phase in the interval $[-180^\circ,\,180^\circ]$ but we also employ other $360^\circ$ intervals to facilitate the presentation of results. In particular, we often employ the phase interval $[-60^\circ,\,300^\circ]$ in order to illustrate fluctuations of the phase about the level of $180^\circ$, which would otherwise appear as $180^\circ$ jumps. In addition, we employ the unwrapped phase lag to better characterise phase dynamics. Results for the  mean  values of the phase difference are typically reported in degrees, whereas instantaneous values of the phase difference are presented in normalised form,  $\phi^*(\tau)=\phi(\tau)/360^\circ$.

\section{Results - force measurements in both reference frames}
\subsection{Direct measurements}\label{sec:direct}
\begin{figure}[t]
\begin{center}
\includegraphics[width=0.635\textwidth]{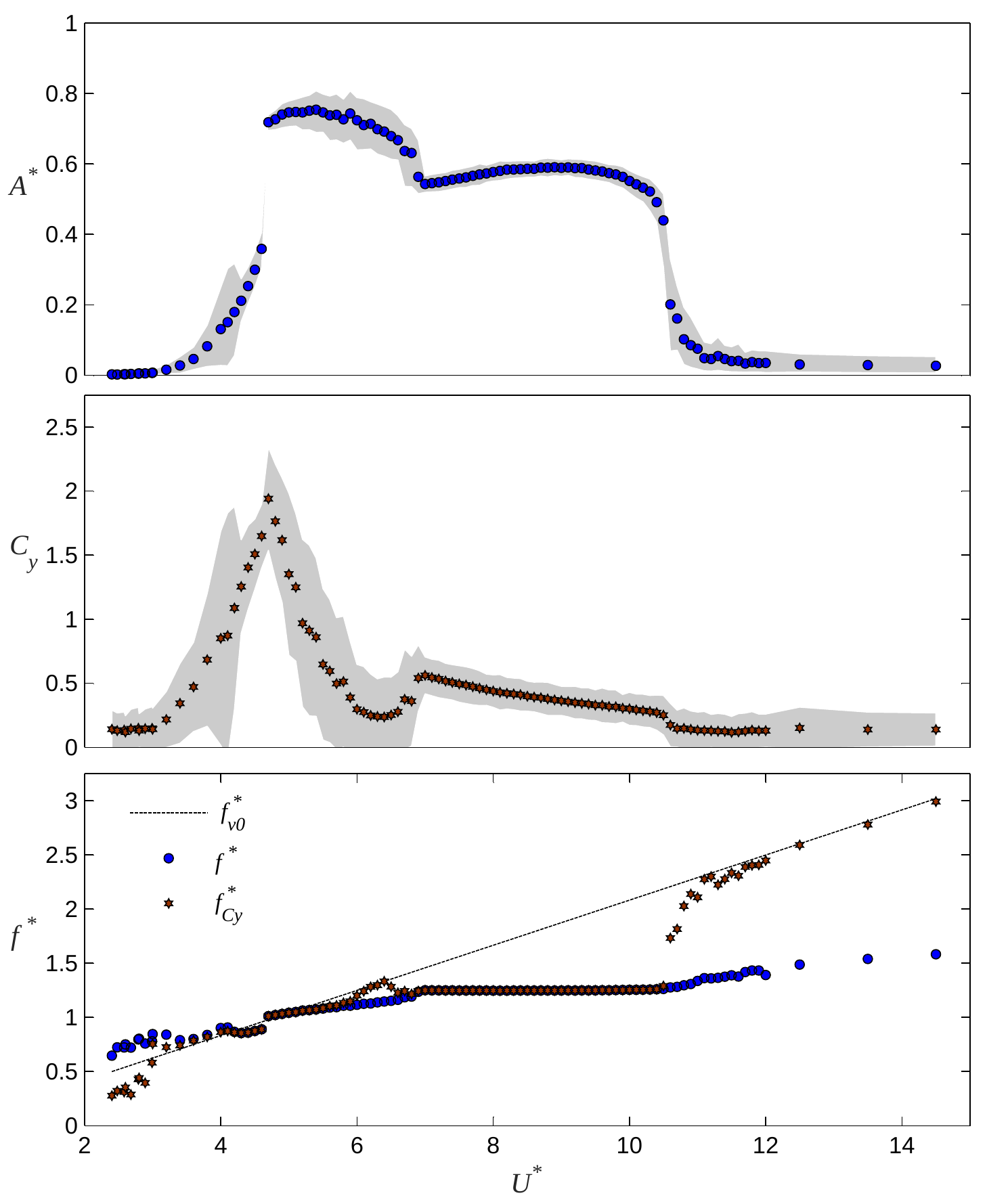}
\end{center}
\caption{Variation of the normalised mean amplitude, $A^*$, the transverse force coefficient, $C_y$, and frequency, $f^*$, of cylinder response and frequency of the transverse force, $f^*_{C_y}$, with reduced velocity, $U^*$; $m^*=3.00; \zeta=0.0035$. The dashed line indicates the normalised frequency of vortex shedding from a stationary cylinder $f^*_{v0}$ assuming a constant Strouhal number of 0.208. Amplitudes of vibration are normalised with the cylinder diameter and frequencies with the natural frequency in still water.} \label{fig:response}
\end{figure}

Figure~\ref{fig:response} shows the variations of the normalised amplitude of  response $A^*$, the transverse force coefficient $C_y$, the normalised frequency of cylinder oscillation $f^*$, and the normalised frequency of the transverse force $f^*_{C_y}$, as functions of the reduced velocity $U^*$. It should be pointed out that reported values denote  mean values of the corresponding instantaneous properties, which were obtained through the Hilbert transform as described in Sect.~\ref{sec:processing}, averaged over the duration of each experimental run.
The variation of $A^*$ with $U^*$ is consistent with previous experiments at low values of $m^*\zeta$ \citep[see, e.g., ][]{Khalak96,Khalak97a,Govardhan00,Brankovich06,Zhao2014a}. Using the terminology in these previous works, we have identified from the  observed variation of the response amplitude the initial, upper, lower, and desynchronisation branches as indicated in Table \ref{table:branches}. 
Two more regions at the lower and upper ends of the reduced velocity range where response amplitudes are very low  have also been included for completeness. In addition, a `bistable region' that marks the transition between the upper and lower branches has also been included, as this will be discussed separately. The plot includes the envelope bracketing the  top  and bottom  5\% levels of vibration amplitudes recorded at each reduced velocity. The width of the envelope shows that amplitude modulations are pronounced in the middle of the initial branch $(3.7<U^*<4.2)$  and in the second half of the upper branch $(5.7<U^*<6.8)$  comparatively to the remaining regions. 

\begin{table}
	\centering
	\small\addtolength{\tabcolsep}{15pt}
\begin{tabular}{lll} \hline
	Branch 	& Range & Amplitude \\ \hline
	-- & $2.4\leqslant U^*<3.2$ & very low \\
	initial & $3.2\leqslant U^*<4.7$ & increasing \\
	upper & $4.7 \leqslant U^* < 6.7 $ & high  \\
	bistable & $ 6.7 \leqslant U^* < 7.0 $ & intermediate \\
	lower & $7.0 \leqslant U^* < 10.5 $ & moderate   \\
	desynchronisation & $\!\!\!\! 10.5 \leqslant U^* <11.1$ & decreasing \\ 
	-- & $\!\!\!\! 11.1 \leqslant U^* \leqslant14.5$ & very low \\ \hline
\end{tabular} 
		\caption{Response  characteristics obtained from the variation of amplitude with reduced velocity.}\label{table:branches}
\end{table}

The middle plot in figure~\ref{fig:response} shows the variation of the transverse force coefficient $C_y$ with $U^*$. The shaded area indicates the 95\% confidence band of instantaneous amplitudes recorded at each reduced velocity. These were computed using the Hilbert transform. $C_y$ increases within the initial branch and reaches a peak value of 1.94 at the start of the upper branch. Then, $C_y$  decreases reaching a minimum value of 0.24 at $U^*=6.4$ within the upper branch and then increases slightly towards the end of the upper branch. At the start of the lower branch $C_y$ exhibits a local maximum with a value of 0.56 and then decreases gradually to approximately half the latter value at the end of the lower branch. The 95\% confidence band of $C_y$ show that the transverse force displays considerable magnitude modulations. In particular, magnitude modulations are more pronounced in the middle of the initial branch and in the second half of the upper branch similarly to the amplitude of vibration. 

As seen in the bottom plot in figure~\ref{fig:response}, $f^*$  increases monotonically from 0.65 to 1.58 from the lowest to the highest $U^*$ value. 
In the initial, upper, transition, and lower branches the frequency of the transverse force, $f^*_{C_y}$, is almost equal to $f^*$, except for the second half of the upper branch $(5.7<U^*<6.8)$ where $f^*_{C_y}>f^*$. However, it should be noted upfront that the instantaneous phase difference between force and displacement exhibits irregular behaviour  in the latter range as will be explicitly shown further below. 
As a consequence,  mean values of $f^*_{C_y}$, which are obtained by time averaging the instantaneous frequency, may be vague in the second half of the upper branch  and should be interpreted with caution. 
In the lower branch, $f^*$ remains constant at approximately 1.25, which is comparable to 1.275 predicted using the empirical formula (eq. \ref{eq:GWformula}) given by  \citet{Govardhan00}. Overall, the variation of the response frequency with the reduced velocity is consistent with previous experiments at low values of mass--damping \citep{Khalak97a,Govardhan00,Zhao2014a}.

\begin{figure}[t]
\begin{center}
\includegraphics[width=0.63\textwidth]{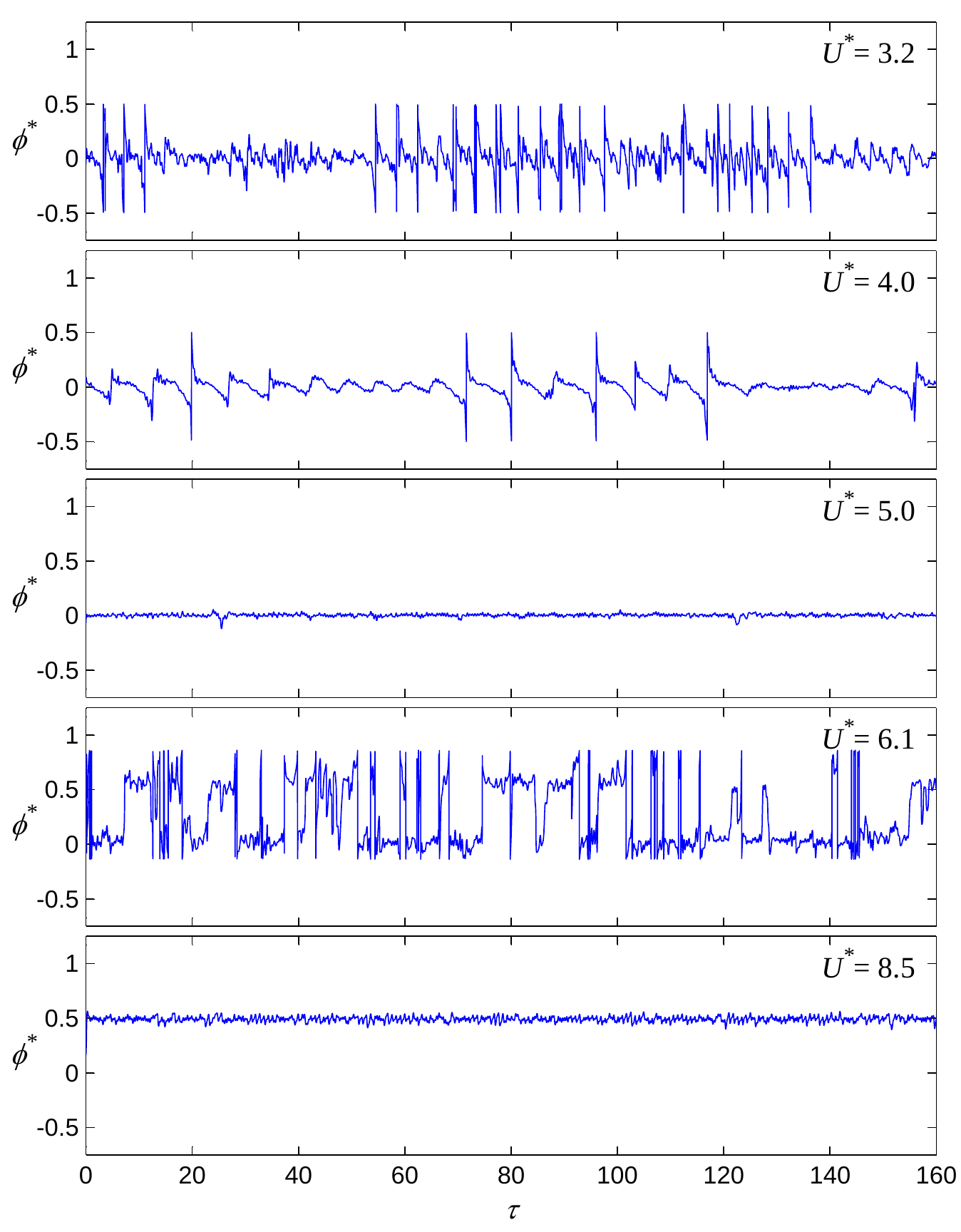}
\end{center}
\caption{Time series of the instantaneous phase difference between transverse force and  displacement at five reduced velocities; initial branch: $U^*=$ 3.2 and 4.0; upper branch:  $U^*=$ 5.0 and 6.1; lower branch: $U^*=8.5$. The normalised phase, $\phi^*=\phi/360^\circ$, is wrapped in the interval $[-1, 1]$ except for $U^*=6.1$ where the phase is  wrapped in the interval $[-1/6, 5/6]$ for improved visualisation.}\label{fig:direct_phase_overall}
\end{figure}

In the following, we focus on the phase dynamics by presenting time series of the instantaneous phase difference between the transverse force and the displacement. At the lower and higher ends of the reduced velocity  range, the instantaneous phase displays highly random fluctuations due to the absence of synchronisation, which are not shown for economy of space. Figure \ref{fig:direct_phase_overall} shows the normalised phase $\phi^*$ at  five different reduced velocities within the synchronisation range. At the start of the initial branch ($U^*=3.2$), the phase displays some random fluctuations about the zero level  whereas in the middle of the initial branch the fluctuations about the mean level become  quasi-periodic ($U^*=4.0$). This quasi-periodicity subsides towards the end of the initial branch where the phase stabilises at $\phi^*\approx0$. At  the start of the upper branch ($U^*=5.0$) the phase remains stable, i.e.\ the transition from the initial branch to the upper branch does not involve any marked change in the phase dynamics. In the middle of the upper branch ($U^*=6.1$), $\phi^*$ intermittently switches between levels of approximately 0 and 0.5 at random instants. 
The random phase jumps persist in the second half  of the upper branch $(5.7<U^*<6.8)$. This change in the phase dynamics across the upper branch is suggestive of a change in character of the response, and that labelling this range of $U^*$ as a single branch may not tell the entire story. At $U^*=8.5$, the phase difference remains remarkably stable around $\phi^*\approx0.5$ ($\phi\approx180^\circ$), which is typical throughout the lower branch. 

\begin{figure}[t]
\begin{center}
\includegraphics[width=0.63\textwidth]{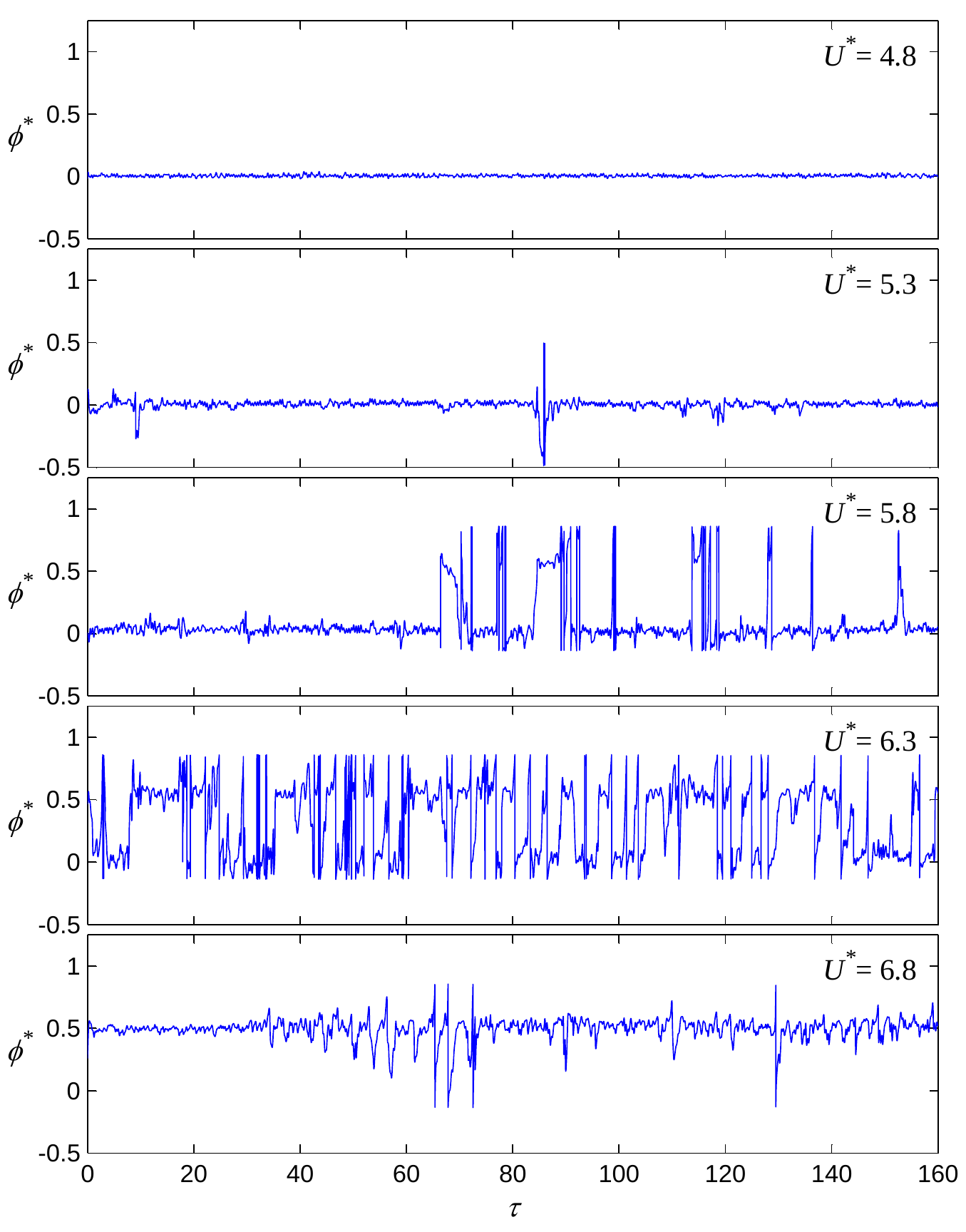}
\end{center}
\caption{Time series of the instantaneous phase difference between transverse force and displacement at different reduced velocities in the upper branch: $U^*=$ 4.8, 5.3, 5.8, 6.3, and 6.8. The normalised phase,    $\phi^*=\phi/360^\circ$, is  wrapped in the interval $[-1, 1]$ except for $U^*=5.8$ and 6.3  where the phase is wrapped in the interval $[-1/6, 5/6]$ for improved visualisation.}\label{fig:phase_direct_upper_wrapped}
\end{figure}

Figure \ref{fig:phase_direct_upper_wrapped} shows time series of the phase difference exclusively in the upper branch.  At the beginning of the upper branch $(U^*=4.8)$, the instantaneous phase remains stable at $\phi^*\approx0$. With increasing the reduced velocity,  sudden phase jumps to the $\phi^*\approx0.5$  level  become more and more frequent ($U^*=$5.3, 5.8, and 6.3). At the end of the upper branch $(U^*=6.8)$   the instantaneous phase stabilises at $\phi^*\approx0.5$  barring few occasional excursions to the zero level. The random phase dynamics observed near the middle of the upper branch may reflect the chaotic character of vortex-induced vibration discovered by \citet{Zhao2014b}. However, a question that arises is whether  sudden jumps in the phase difference are caused by some drastic changes in the flow dynamics, such as instantaneous  swaps in the phasing of vortex shedding induced by sudden changes in the mode of vortex shedding. Most likely, such physical changes require finite time to take place, e.g.\ shed vortices cannot  suddenly swap positions on different sides of the wake. Thus, it seems more likely that small variations in the phasing of vortex shedding are responsible for the sudden  jumps in instantaneous $\phi$. \citet{Leontini2006b}  nicely illustrated how small changes in the pressure balance between the effective stagnation point and low pressure region due to vortices in the formation region can result in $180^\circ$ changes in  $\phi$. 

\begin{figure}
\begin{center}
\includegraphics[width=0.63\textwidth]{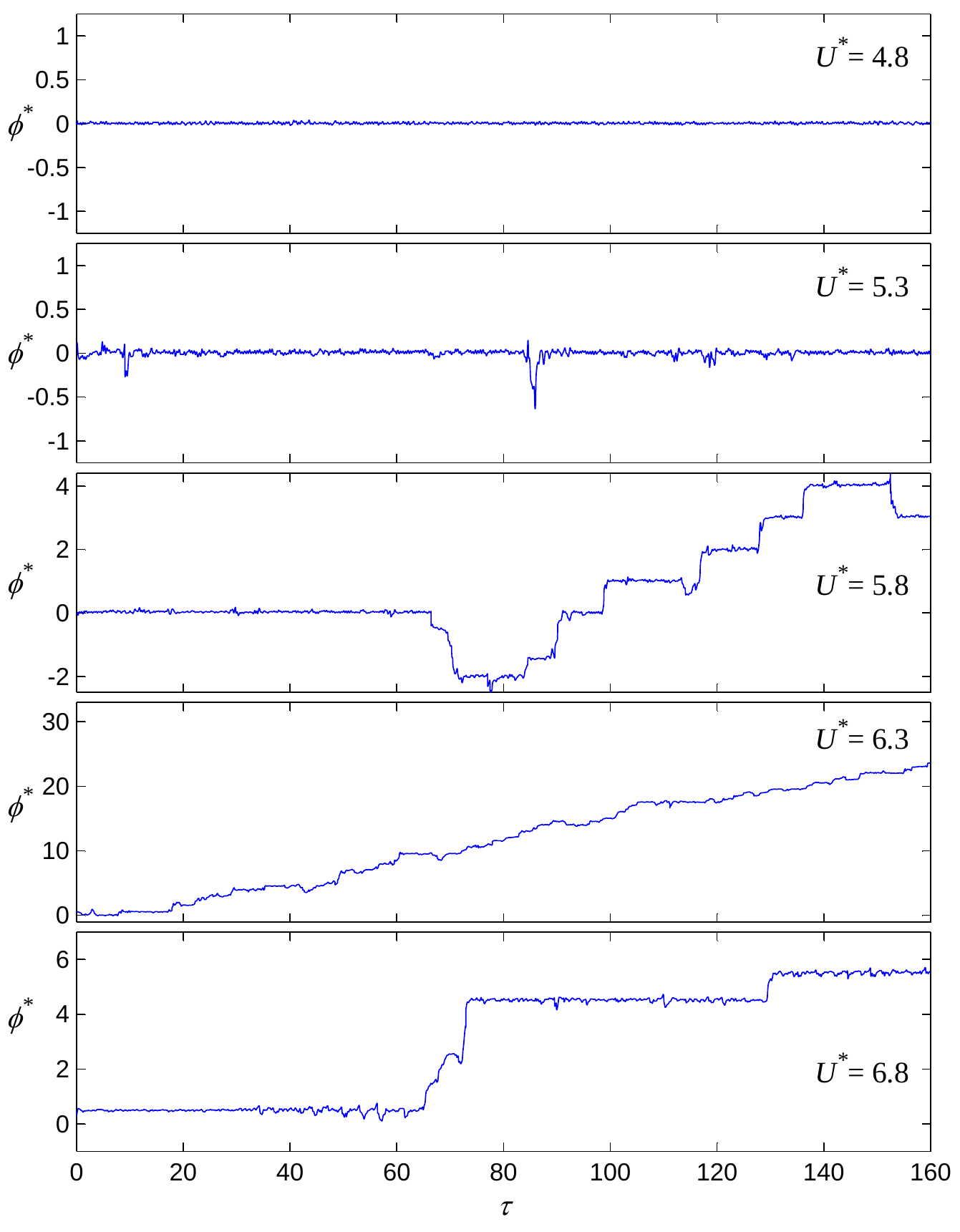}
\end{center}
\caption{Time series of the unwrapped phase difference between  transverse force and  displacement at different reduced velocities in the upper branch: $U^*=$ 4.8, 5.3, 5.8, 6.3, and 6.8. }\label{fig:direct_phase_upper_unwrapped}
\end{figure}

Figure~\ref{fig:direct_phase_upper_unwrapped} shows time series of the unwrapped phase for the same reduced velocities  in the upper branch as in figure~\ref{fig:phase_direct_upper_wrapped}. Now, it can be clearly seen that, e.g.\ at $U^*=5.8$, the unwrapped phased  displays sudden jumps by $\Delta\phi^*=\pm n$, where $n$ is a small integer number, typically 1 or 2. This corresponds to periods of almost constant phase $\phi^*\approx0$ and phase slips in-between. This  indicates intermittent phase dynamics, a feature which typically marks the transition in the border of synchronisation \citep{Pikovsky2001}. We may argue that $\phi$ is effectively constant at $U^*=5.8$ as shown in figure \ref{fig:direct_phase_overall}, as the small epochs of time where $\phi$ deviates from this are associated with a jump of $\pm(0.5+n)$. When  $U^*$ increases to $6.3$,  $\phi^*$ drifts continuously resulting in an unbounded growth.  The drifting behaviour prevails in the second half of the upper branch, i.e.\ over the range $6.0<U^*<6.8$. In the context of nonlinear dynamics, the continuous drift in phase  corresponds to the loss of phase locking, i.e.\ loss of synchronisation \citep{Pikovsky2001}. 
Once the bistable region is entered at  $U^*=6.8$,  periods of effectively constant phase re-emerge with phase slips separating  them. As shown earlier, after the upper$\leftrightarrow$lower transition is completed, the phase difference remains very stable at $\phi^*=0.5$, i.e.\ synchronisation is re-established in the lower branch. Although the time series of the unwrapped phase  do provide some complimentary information, the phenomenological loss of phase locking in part of the upper branch cannot be accounted for.

Table~\ref{table:branches2} summarises the main observations on the basis of the instantaneous phase difference between transverse force and displacement. It is interesting to note that the ranges of different phase dynamics do not  correspond  to the response branches identified from the variation of amplitude with reduced velocity shown in table~\ref{table:branches}.  

\begin{table}
	\centering
\begin{tabular}{lll} \hline
	Range & Phase dynamics &  Branch(es)\\ \hline 
	$3.2\leqslant U^*<3.8$ & fairly stable with distinct jumps (phase slips) & initial \\
	$3.8\leqslant U^*<4.3$ & quasi-periodic & initial \\
	$4.3 \leqslant U^* < 5.5 $ & stable & initial+upper \\
	$5.5 \leqslant U^* < 6.1 $ & stable with distinct jumps (phase slips) & upper\\
	$6.1 \leqslant U^* < 6.7 $ & drifting & upper\\
	$ 6.7 \leqslant U^* < 7.0 $ & stable with distinct jumps (phase slips) & transition\\
	$7.0 \leqslant U^* < 10.5 $ & very stable & lower\\
	$\!\!\!\! 10.5 \leqslant U^* <11.1$ & random & desynchronisation\\ \hline
\end{tabular} 
	\caption{Response characteristics based on the instantaneous phase difference between transverse force and displacement.}\label{table:branches2}
\end{table}

\begin{figure}
\begin{center}
\includegraphics[width=0.675\textwidth]{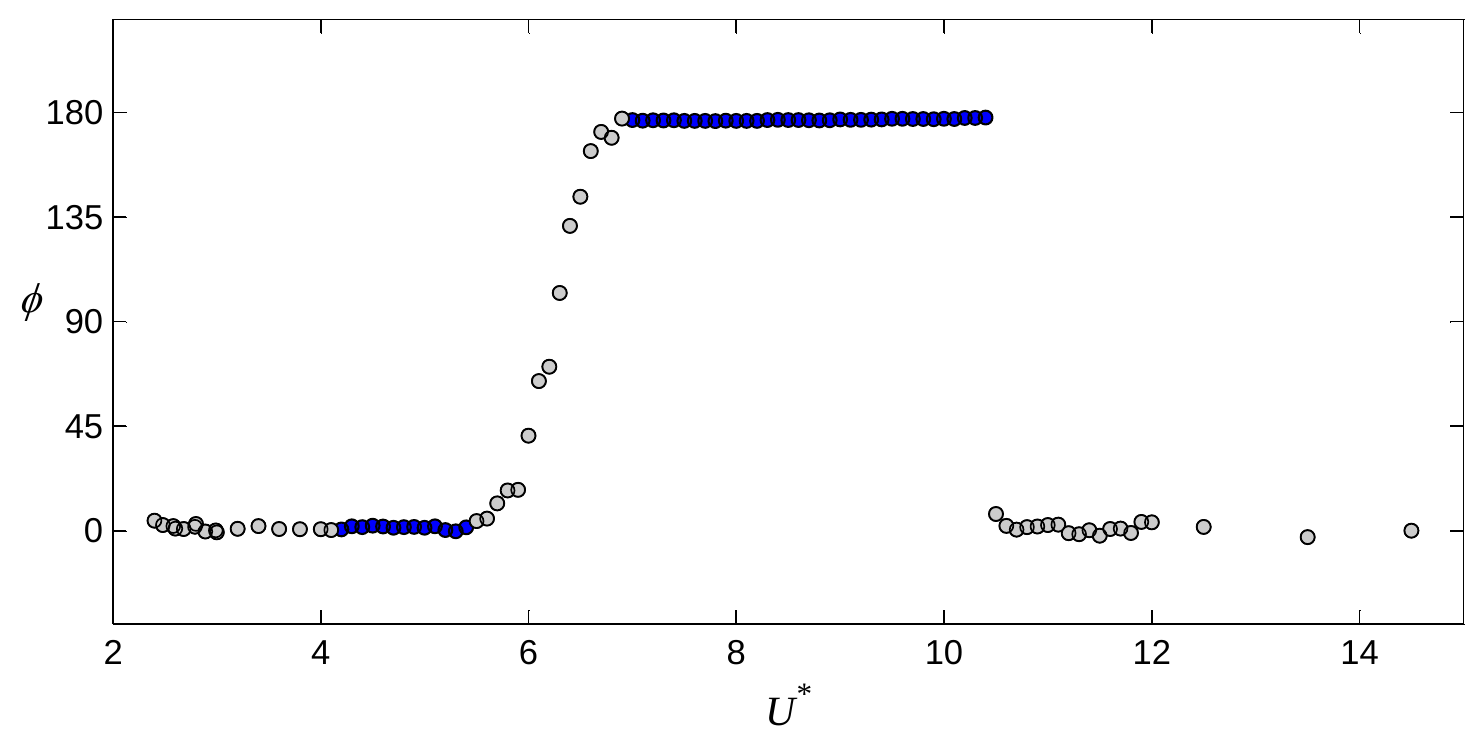}
\end{center}
\caption{Variation of the time-averaged phase lag $\phi$ between transverse force and displacement as a function of  $U^*$. Grey-filled symbols indicate cases for which the instantaneous phase exhibits intermittent jumps and/or continuous drifting and therefore it is not appropriate to characterise the phase dynamics solely by a single time-averaged value.}\label{fig:direct_phase_ave}
\end{figure}

On closing this section,  the variation of the time-averaged phase difference between force and displacement as a function of the reduced velocity is presented in figure~\ref{fig:direct_phase_ave}. The mean phase difference remains slightly above zero  from the initial branch up to the middle of the upper branch where it rapidly changes by approximately  $180^\circ$ within the range $5.5<U^*<6.7$. It then remains just below $180^\circ$ throughout the lower branch. Grey-shaded symbols denote points where the time-averaged value of the phase difference cannot appropriately characterise  the dynamics due to sudden jumps or continuous drifting of the instantaneous phase. This occurs in the lower and upper ends of the reduced velocity range and in the second half of the upper branch as has been  discussed above.    

\subsection{Indirect measurements}\label{sec:indirect}
\begin{figure}
\begin{center}
\includegraphics[width=0.7\textwidth]{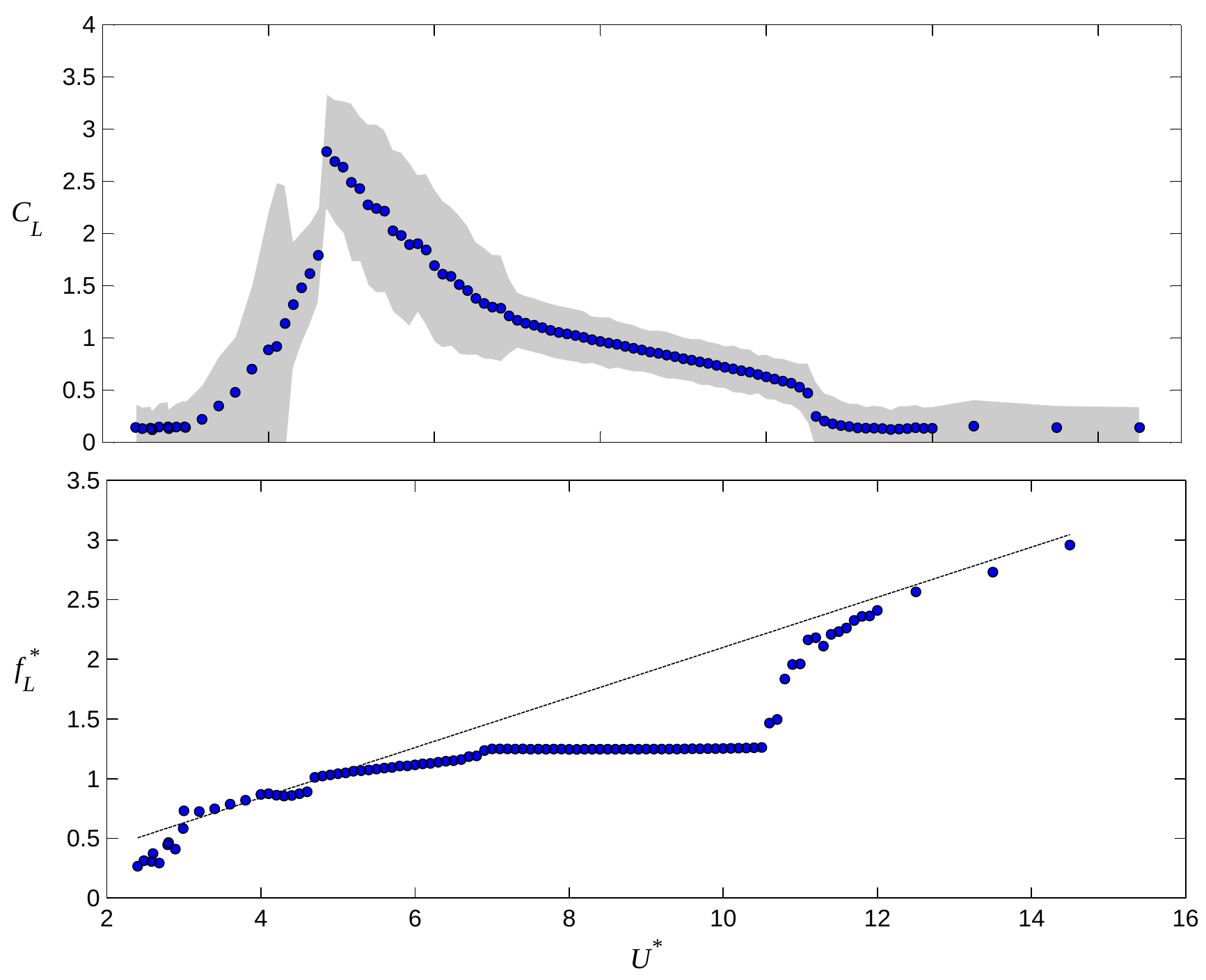}
\end{center}
\caption{Variations of the effective lift coefficient $C_L$ (top plot) and the lift frequency $f^*_{L}$ (bottom plot) with $U^*$. Data represent time-averaged values of the  instantaneous amplitude and frequency of the  effective lift. In the top plot, the shaded area indicates the 95\% confidence band of instantaneous amplitudes recorded at each $U^*$ run.}\label{fig:Lift_vs_Ur}
\end{figure}

In this section, we characterise the hydrodynamics of the freely oscillating cylinder in terms 
of the effective drag and lift components of the total force. Prior to the presentation of the phase dynamics, we present results for the amplitude and frequency of the effective lift and drag components. Figure~\ref{fig:Lift_vs_Ur} shows the variation of the amplitude $C_L$ and  frequency $f^*_{L}$  of the unsteady lift with $U^*$. $C_L$  increases steeply with $U^*$ in the initial branch and jumps to a maximum value of 2.8 at the start of the upper branch. Subsequently, $C_L$ decreases rather steeply over the entire upper branch. The transition from the upper to the lower branch is nearly continuous but is marked by a weak upward kink and a notable change in the rate of decrease of $C_L$ with $U^*$. The end of the lower branch is marked by a sudden drop in magnitude.  The $C_L$ plot includes the envelope of the 95\% confidence band of amplitudes recorded at each reduced velocity. Clearly, the envelope indicates that modulations in $C_L$ are pronounced in the region of quasi-periodic response in the initial branch. In contrast, the envelope is narrow at the end of the initial branch. The envelope becomes comparatively wide in the upper branch, which indicates the existence of considerable amplitude modulations. The bottom plot in figure~\ref{fig:Lift_vs_Ur} shows the variation of the lift frequency $f^*_L$.  Contrary to the abnormal variation of the frequency of the transverse force in the second half of the upper branch  (see figure~\ref{fig:response} and corresponding discussion), $f^*_L$  is perfectly synchronised with the cylinder motion. In the upper branch,  $f^*_L$  increases continuously, whereas $f^*_L$ remains constant in the lower branch, as if  it is limited by some factor. Outside the synchronisation region,  $f^*_L$ tends to follow the straight line corresponding to a constant Strouhal number for a fixed cylinder.  

\begin{figure}
\begin{center}
\includegraphics[width=0.7\textwidth]{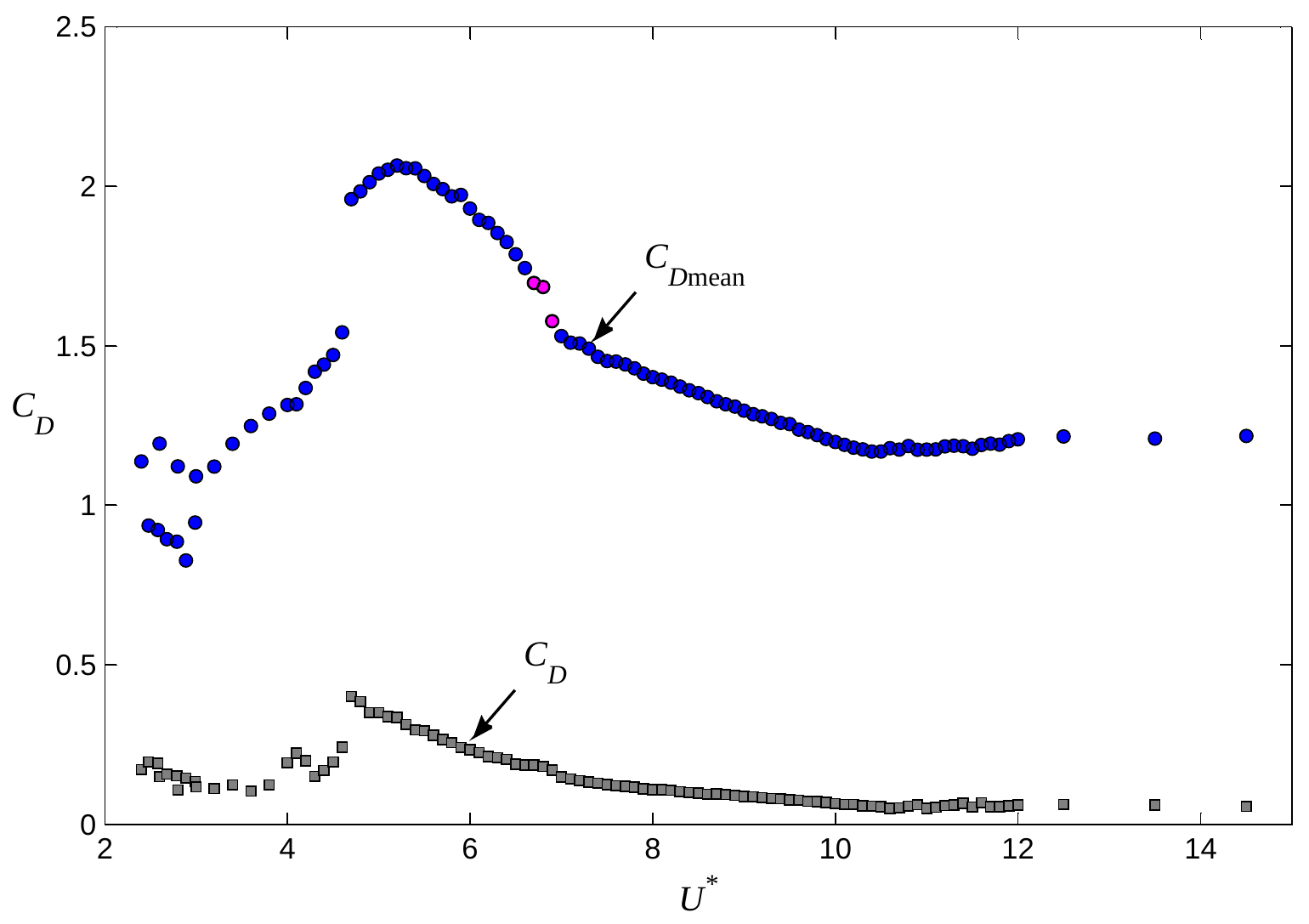}
\end{center}
\caption{(Colour online) Variations of the steady $C_{D\text{mean}}$ and unsteady $C_{D}$    drag coefficients   with $U^*$. Data points marked by magenta colour correspond to the bistable region in order to illustrate the jump in $C_{D\text{mean}}$ accompanying upper$\leftrightarrow$lower branch transition.} \label{fig:Drag_vs_Ur}
\end{figure}

Figure \ref{fig:Drag_vs_Ur} shows the variation of the steady drag coefficient $C_{D\text{mean}}$,  which represents the normalised mean value of the effective drag  averaged over the duration of each experimental run.  $C_{D\text{mean}}$  attains a peak value of approximately 2.07 at $U^*=5.3$, approximately where  the peak response amplitude occurs. The results indicate that $C_{D\text{mean}}\propto A^*$ in the upper branch. The mean drag coefficient is remarkably low before the initial branch. The initial$\leftrightarrow$upper and upper$\leftrightarrow$lower transitions between branches   are marked by distinct jumps in $C_{D\text{mean}}$, excluding data points in the bistable region where each stable state can be characterised by distinct $C_{D\text{mean}}$ values.  In the lower branch $C_{D\text{mean}}$  decreases gradually with $U^*$ to a constant level, which corresponds to approximately the conventional drag coefficient for a non-oscillating cylinder at  corresponding Reynolds numbers. The amplitude of the unsteady drag component, or the unsteady drag coefficient $C_{D}$,  follows a similar trend as a function of $U^*$ as the unsteady lift coefficient $C_L$  (cf.\ figure~\ref{fig:Lift_vs_Ur}).  $C_{D}$ attains a peak value of 0.40 at the start of the upper branch and thereafter  decreases gradually with $U^*$.  

\begin{figure}[t]
\begin{center}
\includegraphics[width=0.62\textwidth]{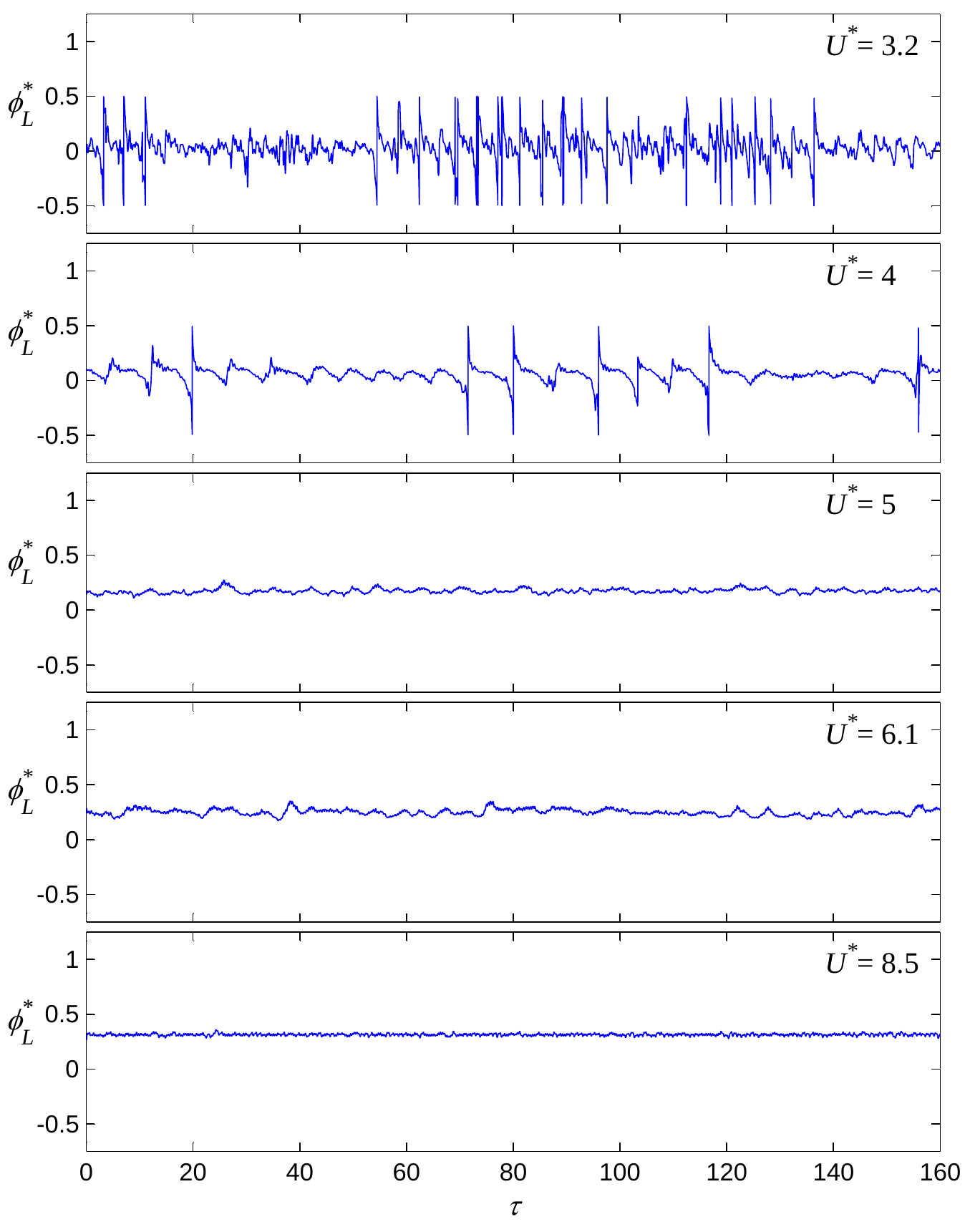}
\end{center}
\caption{Time series of the instantaneous phase difference between the lift component and the displacement at different reduced velocities; initial branch: $U^*=$ 3.2 and 4.0; upper branch:  $U^*=$ 5.0 and 6.1; lower branch: $U^*=8.5$. The normalised phase is  wrapped in the interval $[-0.5, 0.5]$. In these cases, the unwrapped phase is virtually the same except for  $U^*=3.2$ and 4 in which cases there exist some phase jumps. }\label{fig:phase_lift_overall}
\end{figure}

\begin{figure}[t]
\begin{center}
\includegraphics[width=0.62\textwidth]{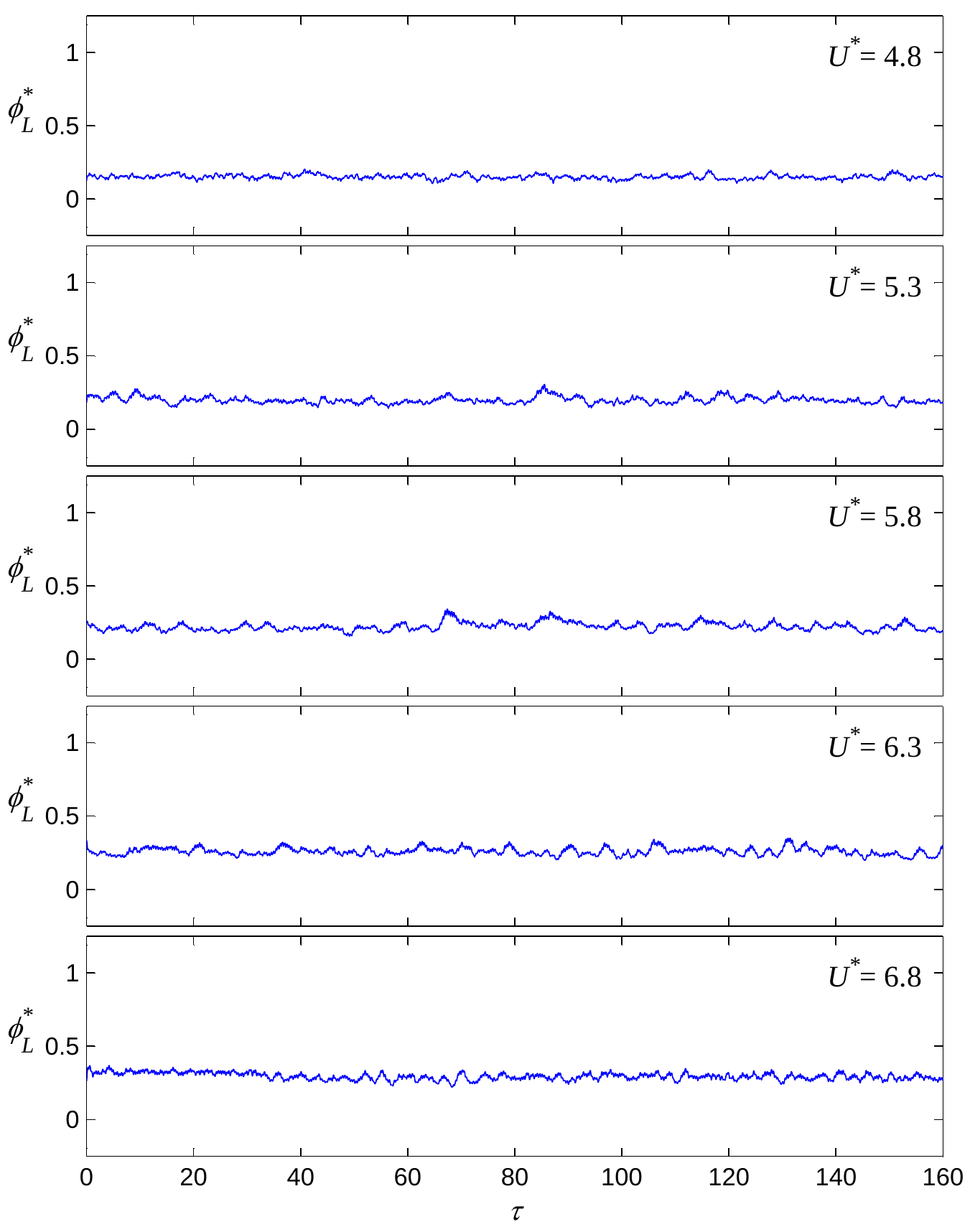}
\end{center}
\caption{Time series of the instantaneous phase difference between the lift component and the displacement at different reduced velocities in the upper branch: $U^*=$ 4.8, 5.3, 5.8, 6.3, and 6.8. The wrapped and unwrapped phases are virtually the same in all cases.}\label{fig:phase_lift_upper}
\end{figure}

The time series of the indirectly-measured lift component were also processed using the Hilbert transform. Figure~\ref{fig:phase_lift_overall} shows the instantaneous phase difference of the lift with respect to the cylinder displacement, $\phi^*_L$ (normalised), at different reduced velocities. At $U^*=$ 3.2 and 4, the dynamics is similar as that of the phase of the transverse force, $\phi^*$  (cf.\ figure~\ref{fig:direct_phase_overall}). This is attributable to the fact that $F_L(t)\approx F_y(t)$ at very small effective angles of attack. At $U^*=$ 5 and 6.1, $\phi^*_L$ remains stable  in contrast to the time traces of $\phi^*$, which display random intermittent jumps at the same reduced velocity (cf.\ figure~\ref{fig:phase_direct_upper_wrapped}). In fact, $\phi^*_L$ remains very stable throughout the upper branch as shown in more detail  in figure~\ref{fig:phase_lift_upper}. This stands in sharp contrast to the existence of sudden jumps and/or continuous drifts of $\phi^*$ in the upper branch (cf.  figure~\ref{fig:direct_phase_upper_unwrapped}). It should be noted here that the time series of the unwrapped phase of the lift  do not differ from the time series of the wrapped phase in the entire upper branch. Furthermore,  $\phi^*_L$ also remains very stable across the entire lower branch  as shown, for example, at $U^*=8.5$ in figure~\ref{fig:phase_lift_overall}. The stable phase difference between lift and displacement is indicative of synchronisation throughout the entire upper and lower branch $(4.2\leqslant U^* \leqslant10.5)$. 

\begin{figure}
\begin{center}
\includegraphics[width=0.7\textwidth]{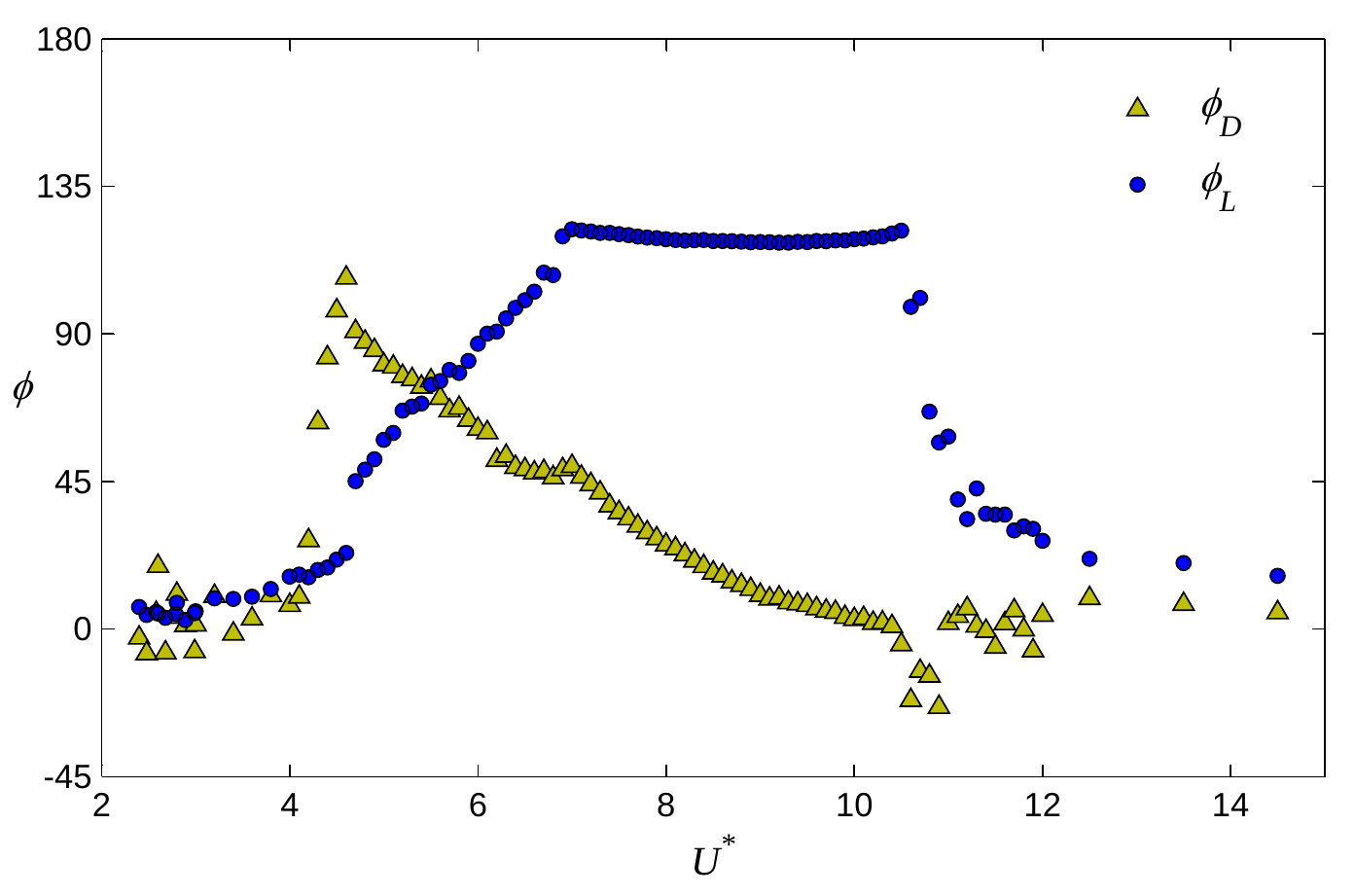}
\end{center}
\caption{Variations of mean phase differences between drag and relative velocity, $\phi_D$, and between lift and displacement, $\phi_L$, with reduced velocity, $U^*$. Mean values are given in degrees.}\label{fig:phase_indirect}
\end{figure}

Figure~\ref{fig:phase_indirect} shows the mean values of $\phi_L$ and $\phi_D$ as functions of $U^*$. Although  data points are included for the entire range of reduced velocities examined, it should be remembered that mean  values of $\phi_L$ are accurate within the range $3.2\leqslant U^*<10.5$ whereas mean values of $\phi_D$  are fairly accurate within a narrower range $4.3\leqslant U^*<10.1$.   The mean phase of the lift with respect to the displacement, $\phi_L$, increases slowly in the initial branch from nearly zero. The transition from the end of initial branch to the start of the upper branch is marked by a distinct jump from $\phi_L=23^\circ$ to $45^\circ$, which nearly corresponds to a doubling in the phase difference. In the upper branch and in the transition region, $\phi_L$ increases almost linearly with $U^*$.  The transition from the end of the upper branch to the start of the  lower branch, excluding three data points in the transition region, involves a second distinct jump from $\phi_L=103^\circ$ to $122^\circ$. In the lower branch, $\phi_L$ remains remarkably constant.  

\begin{figure}
\begin{center}
\includegraphics[width=0.74\textwidth]{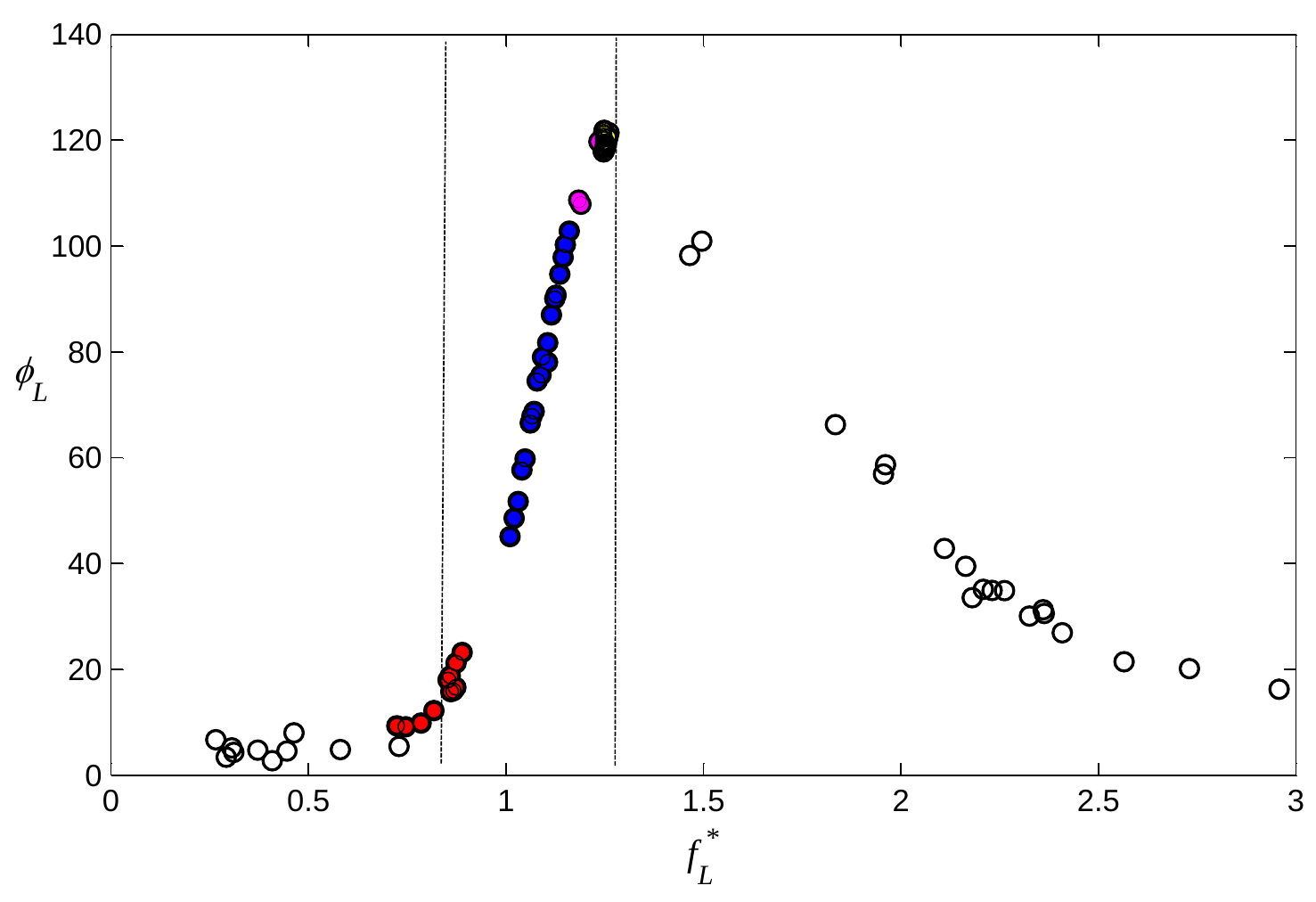}
\end{center}
\caption{(Colour online) Variation of the lift phase, $\phi_L$, with the lift frequency, $f_L^*$. Filled colours denote different response branches: initial (red), upper (blue), transition (magenta), lower (yellow). Note that data points in the lower branch cannot be discerned as they fall on top of each other. Vertical dashed  lines mark the borders of the synchronisation region. }\label{fig:phase_lift_freq}
\end{figure}

From non-linear dynamical systems it is known that frequency  and phase of an oscillator are closely related since  the phase typically changes by $180^\circ$ as the forcing frequency varies across the synchronisation range  \citep{Pikovsky2001}. Indeed, we have actually seen this in figure~\ref{fig:direct_phase_ave}. However, this change takes place over a range of reduced velocities where  the instantaneous phase difference between $F_y(t)$ and $y(t)$ becomes unbounded due to continuous drifting. Therefore, the mean phase $\phi$ takes intermediate values between $0^\circ$ to $180^\circ$,  which are vague; the dynamics become irregular and the  phase of the transverse force with respect to the cylinder oscillation  cannot be unambiguously correlated with the synchronisation frequency within the former range.  
In contrast, the  phase difference between lift and displacement $\phi_L$ varies in an orderly fashion with the lift frequency as shown in figure~\ref{fig:phase_lift_freq}. Prior to the onset of  synchronisation at low $f_L^*$ values $\phi_L$ remains at a low level close to zero. For $f_L^*$ values above the synchronisation region , $\phi_L$ decreases  smoothly towards the zero level. Data points in the synchronisation range, whose borders have been marked by vertical lines on the plot, have a linear correlation coefficient of 0.9889, which indicates a strong linear relationship between $f_L$ and $\phi_L$.  
 In the lower branch where $f_L^*$ remains almost constant, $\phi_L$ also remains almost constant  so that data points fall on top of each other (agglomerated data points at maximum $\phi_L$ values). Another point to note is that  transitions from the initial to the upper branch as well as from the upper to the lower branch are clearly marked by simultaneous jumps in both frequency and phase of the lift (excluding points in the bistable region). 

The variation of  $\phi_L$ as a function of  $f_L^*$ shown in figure~\ref{fig:phase_lift_freq} suggests that  the lift phase unequivocally depends  on the lift frequency, which is  the same as the frequency of vortex shedding in the wake during synchronisation. This  relationship between the lift phase and the frequency of vortex shedding has been previously examined in detail by \cite{Konstantinidis11}. 
They showed that  a direct relationship exists between the phase of the lift and the timing of vortex shedding for the case of synchronisation of the cylinder wake to external forcing by means of periodic perturbations in the velocity of the free stream. Furthermore, \cite{Konstantinidis2016} showed how the kinematics of a cylinder oscillating transversely to a free stream can be  analysed in terms of the relative velocity of the cylinder and the free stream. Therefore, it is reasonable to assume that the phase of the lift component on a cylinder oscillating transversely to a free stream directly reflects the vortex dynamics in the wake, which has been the founding hypothesis for the present study.

\begin{figure}
\begin{center}
\includegraphics[width=0.74\textwidth]{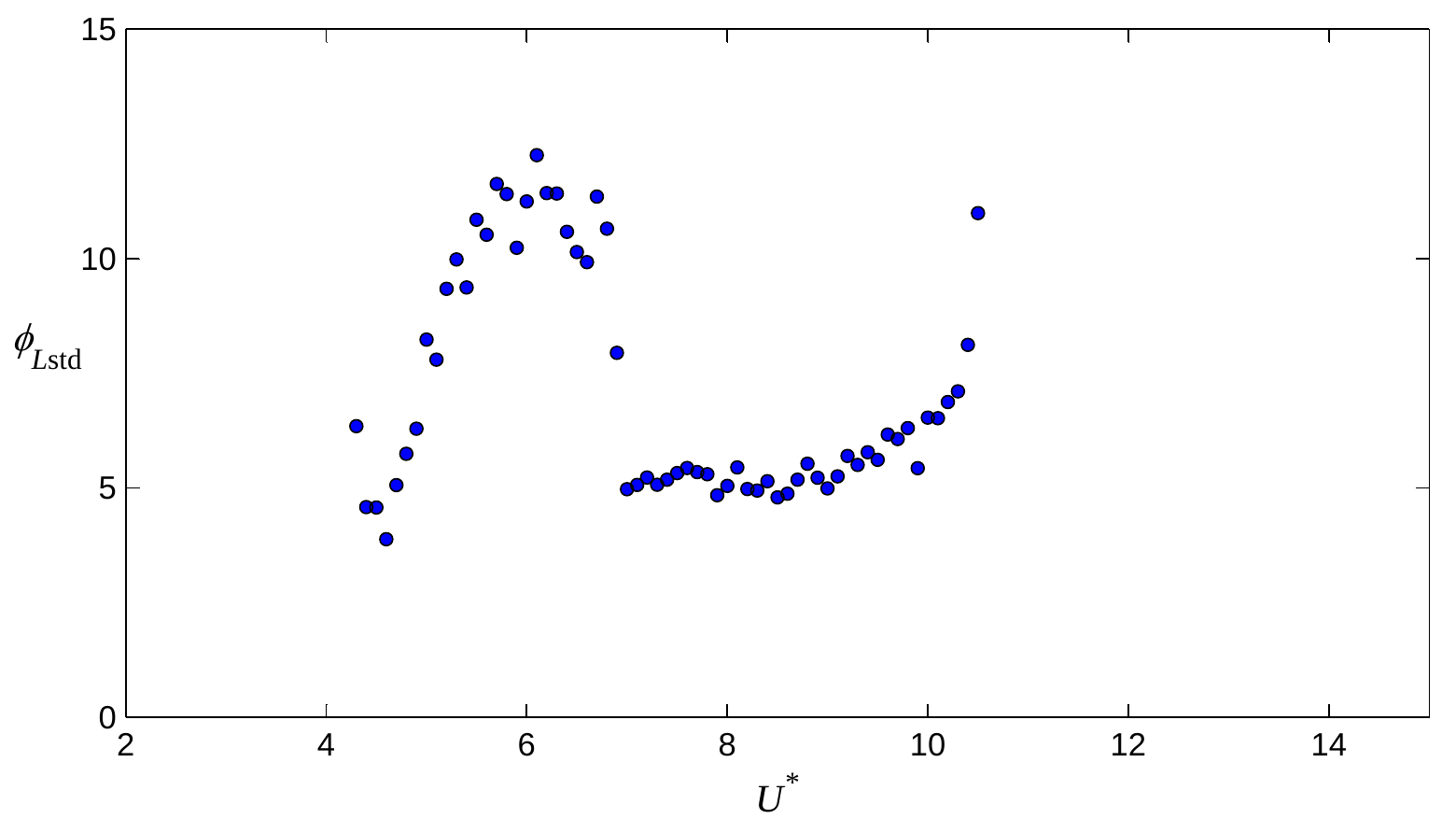}
\end{center}
\caption{The variation of the standard deviation of the lift phase, $\phi_{L\text{std}}$, with reduced velocity, $U^*$.}\label{fig:phase_lift_std}
\end{figure}

As discussed earlier, the phase dynamics of lift exhibits  a notable change in the second half of the upper branch compared to the first half. This change can also be illustrated by another metric: the variation of the standard deviation of the instantaneous $\phi_{L}$, denoted as $\phi_{L\text{std}}$. The variation of $\phi_{L\text{std}}$ as a function of $U^*$ is shown in figure~\ref{fig:phase_lift_std}.  Data points cover the last part of the initial branch and the  entirety of the upper and lower branches where strong phase-locking occurs. For $U^*$  values in the first half of the initial branch,  $\phi_{L\text{std}}$ is very large (exceeds the scale of the plot) but then drops and attains a  minimum value at the end of the initial branch as seen in the plot. Subsequently, $\phi_{L\text{std}}$ increases rapidly in the first half of the upper branch  whereas $\phi_{L\text{std}}$  stays at a high level above $10^\circ$ in the second half of the upper branch. In the lower branch,  $\phi_{L\text{std}}$ drops to  a low constant level of approximately $5^\circ$. Towards the end of the lower branch $\phi_{L\text{std}}$ increases rapidly with $U^*$ and $\phi_{L\text{std}}$ exceeds the scale of the plot  outside the synchronisation region. Therefore, the $\phi_{L\text{std}}$ results  show that  the phase difference between lift and displacement displays relatively pronounced modulations in the second half of the upper branch. Overall, the standard deviation of $\phi_L$ is small in the synchronisation region, which reflects the absence of jumps and/or drifts in $\phi_L$, i.e.\ the  phase difference of the lift component with respect to the cylinder oscillation provides a robust indicator of the dynamics. 

\subsection{Physical function of drag and lift components}
\begin{figure}
\begin{center}
\includegraphics[width=0.72\textwidth]{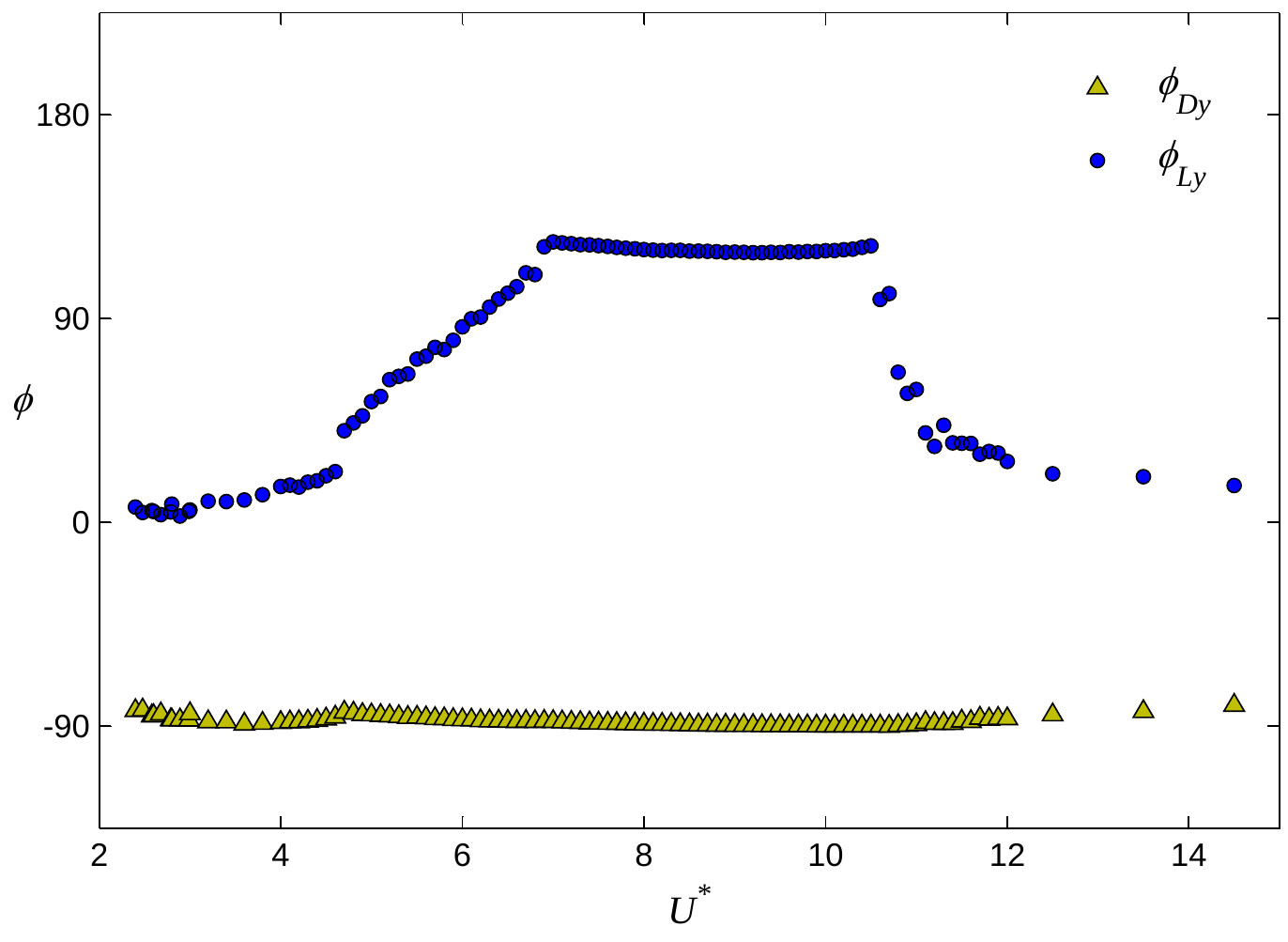}
\end{center}
\caption{Variations of the phase differences $\phi_{Dy}$ and $\phi_{Ly}$ between $F_{Dy}(t)$ and $y(t)$ and between $F_{Ly}(t)$ and $y(t)$, respectively, with reduced velocity, $U^*$. Data show  time-averaged mean values of the instantaneous phase differences obtained using the Hilbert transform of the corresponding signals.}\label{fig:phase_exc_damp}
\end{figure}

To understand the physical function of the drag and lift components in VIV, we can take their instantaneous projections on the direction of cylinder motion ($y$ axis) using the following relationships,
\refstepcounter{equation}
$$
		F_{Dy} = -F_D\sin{a_\text{eff}} \quad\text{and}\quad F_{Ly} = F_L\cos{a_\text{eff}}. \eqno{(\theequation{\mathit{a},\mathit{b}})}\label{eq:projections}
$$
Next, we process the time series of  $F_{Dy}(t)$ and $F_{Ly}(t)$  using the Hilbert transform as we did earlier with the time series of $F_{D}(t)$ and $F_{L}(t)$. However, both $F_{Dy(t)}$ and $F_{Ly}(t)$   are now synchronised with the cylinder motion due to their explicit dependency on $a_\text{eff}$. As a consequence, the differences of the instantaneous phases of $F_{Dy}(t)$ or $F_{Ly}(t)$ and of $y(t)$ are very stable in the initial, upper and lower branches, including the bistable region. Therefore, mean values of the  instantaneous phase differences can be employed with confidence in order to reveal the functions of the drag and lift components. The corresponding phase differences are denoted $\phi_{Dy}$ and $\phi_{Ly}$, respectively, and their variations with $U^*$ are shown in figure~\ref{fig:phase_exc_damp}. Two   observations can be  made immediately. First, $\phi_{Dy}$ remains very close to $-90^\circ$ over the entire $U^*$ range.  These results show that $F_D(t)$ makes a contribution that  always opposes the velocity of the cylinder, predominantly due to the mean drag, i.e.\ the drag  always makes a  contribution to positive damping  of the cylinder vibration. Second, at each reduced velocity $\phi_{Ly}\approx\phi_L$ (cf.\ figure~\ref{fig:phase_indirect}). Thus, $F_L(t)$ induces a component in the direction of oscillation $F_{Ly}(t)$ that is synchronised with the motion resulting in large-amplitude vortex-induced vibration. 

\begin{figure}
\begin{center}
\includegraphics[width=0.72\textwidth]{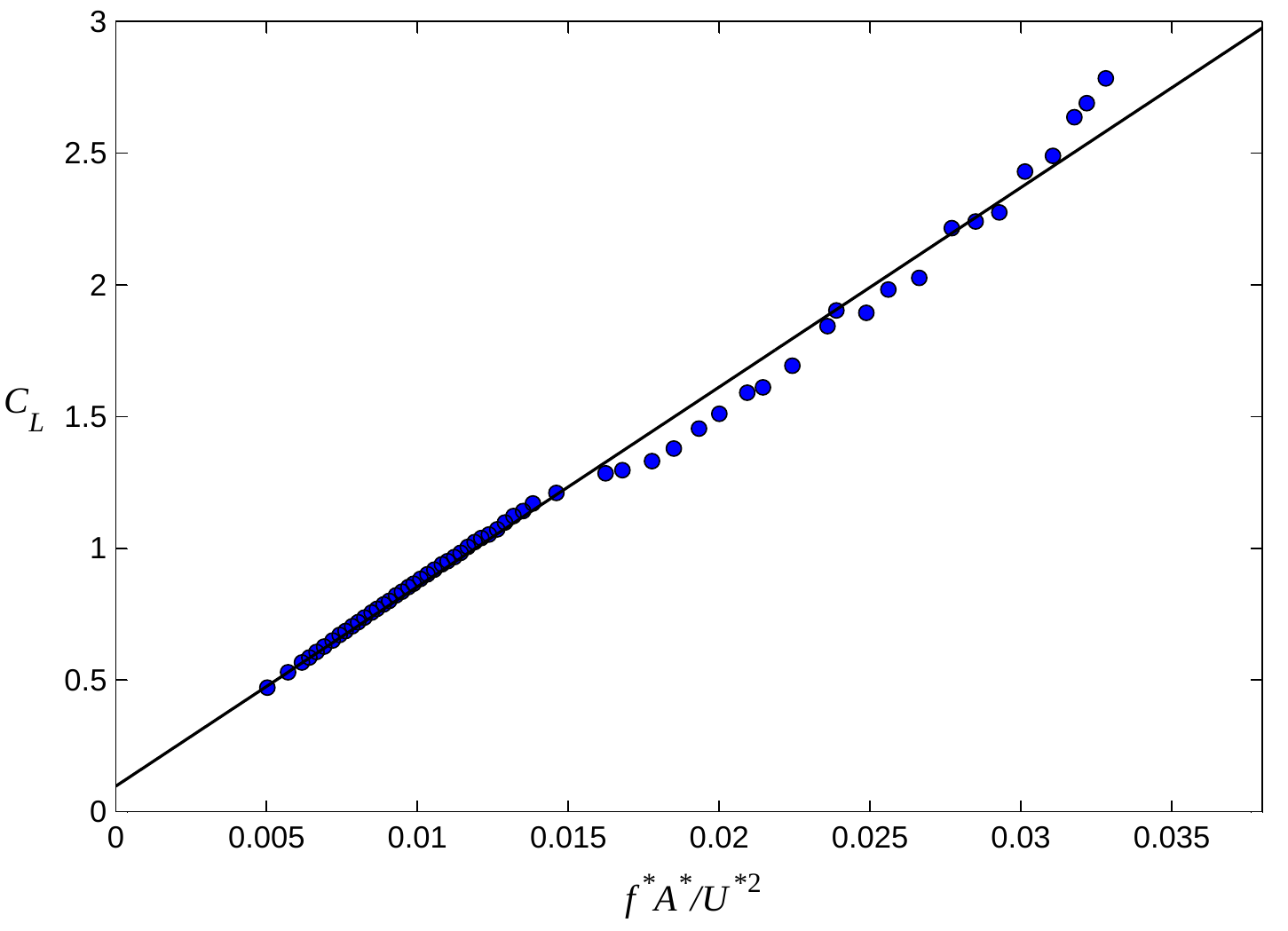}
\end{center}
\caption{Variation of the lift coefficient $C_L$ as a function of  the parameter $f^*A^*/U^{*2}$ within the lower and upper branches. The straight line is a least-squares  fit to the data, which has an R-square value of 0.9998 showing that the relationship is linear.}\label{fig:CLexc}
\end{figure}

Typically, the component of the force in-phase with the cylinder velocity $C_y\sin\phi$ is considered to be the excitation force coefficient in VIV. According to the equation of motion (if the motion is harmonic), $C_y\sin\phi \propto f^*A^*/U^{*2}$, where the proportionality factor is proportional to the damping ratio. For very low values of the damping ratio $\zeta$, the proportionality factor is itself very low. This considerably limits the permissible range of values of the phase angle, i.e.\ $\phi$ has be just above $0^\circ$ or just below $180^\circ$. A very important finding from the present study, as illustrated in figure~\ref{fig:CLexc}, is that the lift coefficient varies linearly with the same scaling factor as does the typical excitation coefficient. A best-linear fit to the data, 
\begin{equation}\label{eq:CLfit}
	C_L = C_{L_0}+ \kappa\,\frac{f^*A^*}{U^{*2}}.
\end{equation}
yields $ C_{L0}=0.1$ and $\kappa=75.8$ with a R-square value of 0.9998, which reveals a highly linear relationship. This is a remarkable result since now there is  no direct  dependency on the phasing of the lift force, but it is embodied   through the variation of $f^*$. The constant factor, $C_{L0}$, corresponds to approximately the lift coefficient of a non-vibrating cylinder at similar Reynolds numbers.

\section{Results - combining indirect forces with harmonic modelling}
\subsection{Modelling of hydrodynamics}
The previous sections have shown that the phase behaviour of the indirect forces is much more stable than the phase of the transverse force. Here, we investigate whether these indirect forces can be combined with a harmonic model of oscillation to further understand VIV.

Assuming harmonic oscillation the lift and drag forces can be written as
\begin{equation}\label{eq:lift_model}
	F_L(t) = \frac{1}{2}\rho U_\infty^2DL C_L\sin{(\omega t+\phi_L}),
\end{equation}
and
\begin{equation}\label{eq:drag_model}
	F_D(t) = \frac{1}{2}\rho U_\infty^2DL \left[ C_{D\text{mean}} + C_D\sin{(2\omega t+\phi_D}) \right],
\end{equation}
respectively, where $\omega=2\pi f$ is the angular frequency of cylinder oscillation in radians/s. Note that the lift fluctuates at the cylinder oscillation frequency whereas  the drag fluctuates at twice the frequency of cylinder oscillation. Further assuming small angles (the maximum measured $\alpha_\text{eff}$ at the start of the upper branch, is approximately $11^\circ$), then $\sin\alpha_\text{eff}\approx\dot{y}/U_\infty$ and $\cos\alpha_\text{eff}\approx1$. This assumption implies that $F_{L_y} \simeq F_L$ and $\phi_{Ly}\approx\phi_L$ - in excellent agreement with the results in figures \ref{fig:phase_indirect} and \ref{fig:phase_exc_damp}. The forces can therefore be written as 
\begin{equation}\label{eq:FLy_approx}
	\frac{F_{Ly}(t)}{\frac{1}{2}\rho U_\infty^2DL} \approx C_L \sin{(\omega t+\phi_L)} + \cdots,
\end{equation}
and
\begin{equation}\label{eq:FDy_approx}
	\frac{F_{Dy}(t)}{\frac{1}{2}\rho U_\infty^2DL} \approx - \left(\frac{\omega A}{U_\infty} \right) \left[C_{D\text{mean}} \sin{\left(\omega t-\frac{\pi}{2}\right)} + \frac{1}{2}C_D \sin{(\omega t+\phi_D)} + \cdots \right],
\end{equation}
where higher order harmonics of the primary frequency of cylinder oscillation have been neglected.

Equation (\ref{eq:FDy_approx}) shows that $F_{Dy}(t)$ comprises two separate contributions from the mean  and the fluctuating drag. As seen in figure \ref{fig:Drag_vs_Ur}, $C_D$ values are considerably less than $C_{D\text{mean}}$ values. Therefore, the sum in the square brackets of equation (\ref{eq:FDy_approx}) is dominated by the first term related to the mean drag. 
So, $\phi_{Dy}$  takes values close to $-90^\circ$ for all values of $U^*$, in excellent agreement with the results shown in figure~\ref{fig:phase_exc_damp}. Thus, the above harmonic hydrodynamical model is consistent with measurements of the mean phases of  lift and drag components and is employed in the following subsections  to exemplify several aspects of VIV.

\subsection{The onset of chaotic oscillations in the upper branch}
\begin{figure}
\begin{center}
\includegraphics[width=0.99\textwidth]{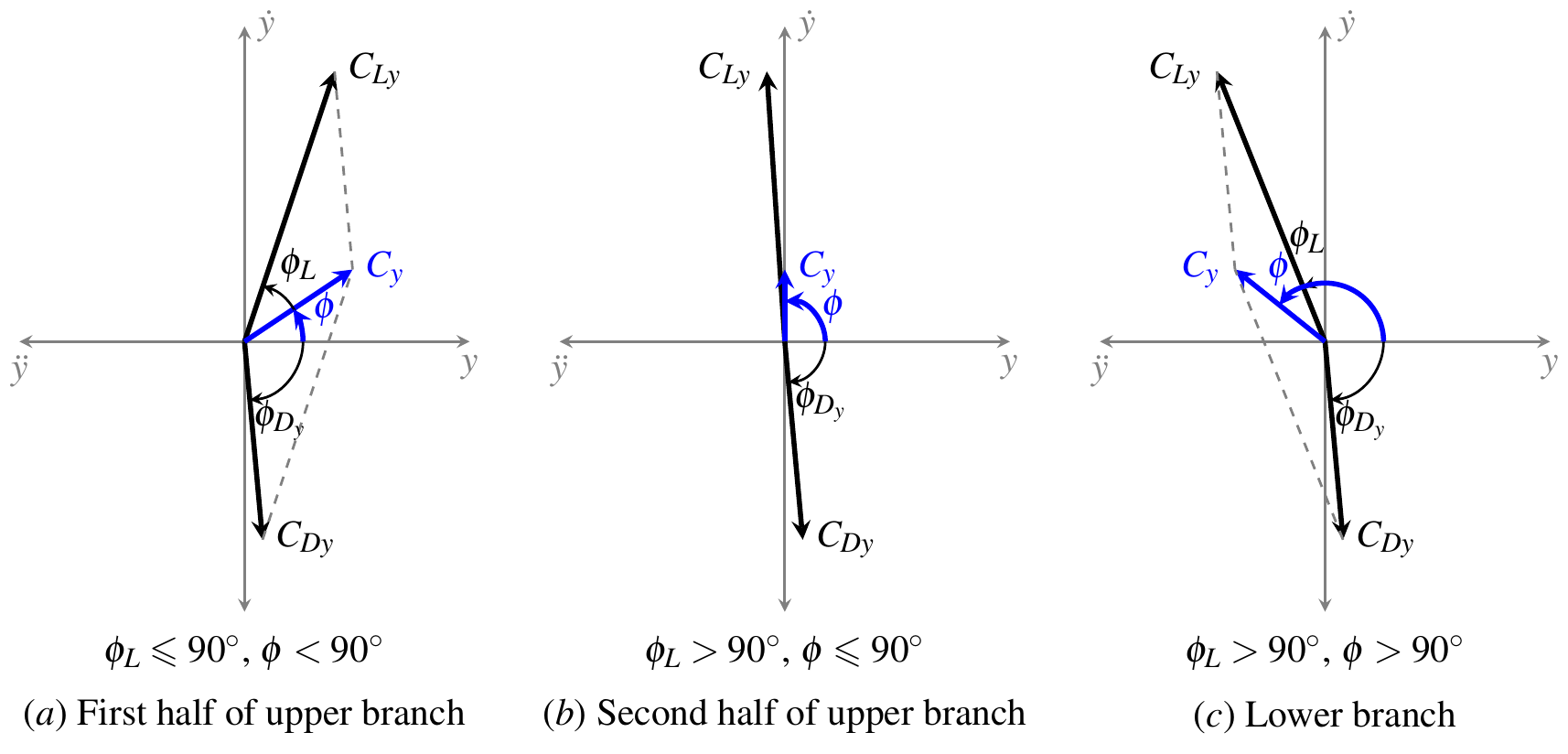}
\end{center}
\caption{(Colour online) Phase diagrams of the drag and lift components of the transverse fluid force in three regions with different dynamics. The magnitude and angle of the phasors are not to scale for better visualisation. }\label{fig:phase_diagram_simple}
\end{figure}

The phenomenology of chaotic dynamics observed in the second half of the upper branch can be understood with the aid of phase diagrams of the $F_{Ly}$ and $F_{Dy}$ components of the hydrodynamic force shown in figure~\ref{fig:phase_diagram_simple}, where phasors (vectors)  represent the magnitude and phase of  each component with respect to the cylinder displacement ($y$ axis) and velocity ($\dot{y}$ axis). The phasor of the resulting transverse force, which is the vectorial sum of the components of drag and lift in the direction of motion,  is depicted in blue colour in figure~\ref{fig:phase_diagram_simple}. 

For reduced velocities in the first half of the upper branch, $\phi_L\leqslant 90^\circ$ and the resulting transverse force has a significant component in-phase with the cylinder velocity (figure~\ref{fig:phase_diagram_simple}{\it a}). As the reduced velocity is increased, $\phi_L$ approaches to $90^\circ$ in the middle of the upper branch. For reduced velocities in the second half of the upper branch, $90^\circ<\phi_L<103^\circ$ and the phasors of  $F_{Ly}$ and $F_{Dy}$ are pointing to approximately opposite directions in the phase diagram (figure~\ref{fig:phase_diagram_simple}{\it b}). As a result, they cancel each other out  leaving only a small component of the transverse force almost in-phase with velocity. 
As the reduced velocity is increased further, $\phi_L$ jumps to $122^\circ$ in the lower branch and the resulting transverse force has a significant component in-phase with the acceleration. 

When the phase difference between the forcing (i.e.\ the hydrodynamic lift) and the response (i.e.\ the cylinder motion) passes through the point where $\phi_L\approx90^\circ$, the hydro-elastic system may be considered to be in a state of `unstable equilibrium'; although the phase difference between forcing and response  $\phi_L$ remains bounded, small perturbations can cause amplitude modulations, i.e.\ perturbations have the largest impact  in the system's response \citep{Pikovsky2001}. Modulations in the amplitude of cylinder oscillation have a feedback effect on the fluid forcing causing the system to behave chaotically breaking the assumption of harmonic oscillation used for our model. This occurs at around $U^*\approx6.1$, where $\phi_L=90.05^\circ$ and  $\phi_{L\text{std}}$ peaks (see figures \ref{fig:phase_indirect} and \ref{fig:phase_lift_std}), which supports our hypothesis that the upper branch consists of a limit cycle which is stable at low $U^*$, but unstable at high $U^*$. This is consistent with the earlier observations of the onset of chaos in the upper branch in \citet{Zhao2014a}.  

\subsubsection{Revealing competing factors through formulae}\label{sec:factors}
\begin{figure}
 \begin{center}
 \includegraphics[width=0.67\textwidth]{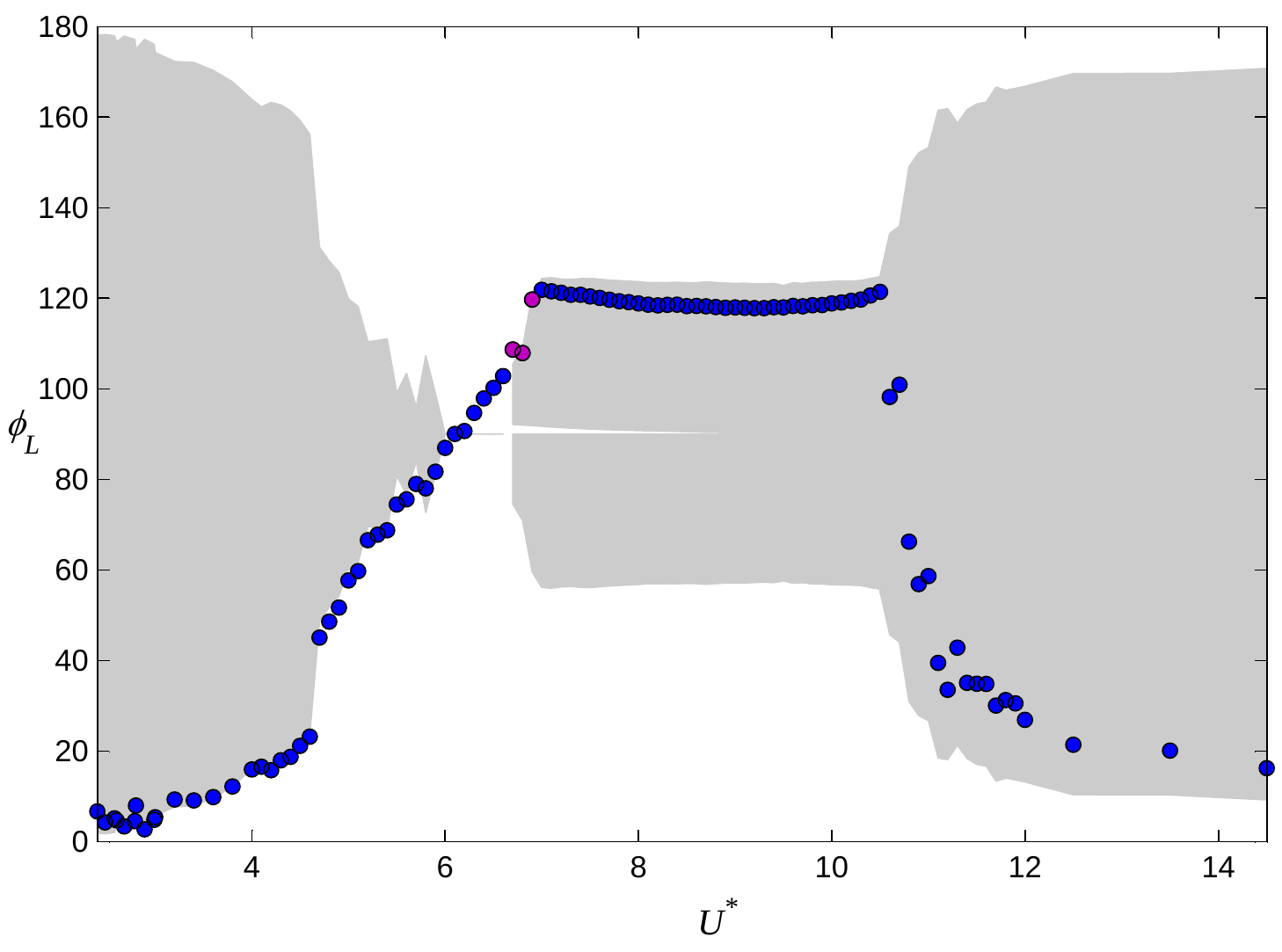}
 \end{center}
 \caption{(Colour online) Permissible range of the phase difference $\phi_L$ as a function of the reduced velocity $U^*$ computed from the inequalities in (\ref{eq:cond3}).  Values of $\phi_L$ are in degrees. Magenta filling indicates points in the bistable range.}\label{fig:phase_range}
 \end{figure}

The dynamics of VIV can be further illustrated through analytical formulae governing the hydrodynamics and the cylinder motion.  
 From equations (\ref{eq:FLy_approx}) and (\ref{eq:FDy_approx}), we can express the component of the transverse force in phase with the velocity of the cylinder as
 \begin{equation}
 	C_y\sin\phi =  C_L\sin\phi_L - \frac{2\pi f^*A^*}{U^*} \left( C_{D\text{mean}}  + \frac{1}{2}C_D\sin\phi_D \right).
 \end{equation}
 Assuming that the cylinder displacement can be approximated, on the average, as sinusoidal $y(t)\approx A\sin{(\omega t)}$, the condition  $C_y\sin\phi>0$, which is required for positive excitation in free vibration (see Introduction), can be reformulated as 
 \begin{equation}\label{eq:cond2}
 	C_L\sin\phi_L > \frac{2\pi f^*A^*}{U^*} \left(C_{D\text{mean}} + \frac{1}{2}C_D\sin\phi_D \right).
 \end{equation}
 The inequalities in (\ref{eq:cond2})  restrict the phase difference between lift and displacement in the range
 \begin{equation}\label{eq:cond3}
 	 \sin^{-1}\left\{ W(A^*,f^*\!/U^*) \right\}  < \phi_L < \frac{\pi}{2} - \sin^{-1}\left\{ W(A^*,f^*\!/U^*) \right\},
 \end{equation}
 where
 \begin{equation}\label{eq:W}
 	 W(A^*,f^*\!/U^*) = \frac{2\pi f^*A^*}{U^*}\left( \frac{C_{D\text{mean}}}{C_L}  + \frac{1}{2}\frac{C_D\sin\phi_D }{C_L}\right).
 \end{equation}
 The term $W(A^*,f^*\!/U^*)$ represents a function of the normalised amplitude and frequency, which generally varies with the reduced velocity.  The permissible range of $\phi_L$ values given by (\ref{eq:cond3}) is plotted along with $\phi_L$ measurements in  figure~\ref{fig:phase_range}. The permissible range of $\phi_L$ values becomes narrower as the reduced velocity is increased in the initial and in the first half of the upper branch $(4.7\leqslant U^* <6.1)$. At the middle of the upper branch, $U^*\approx6.1$, $\phi_L$ is  restricted to be exactly $90^\circ$ in agreement with indirect measurements. As discussed earlier, this point corresponds to the state of unstable equilibrium. In the second half of the upper branch, $W(A^*,f^*\!/U^*)\approx1$ so that $\phi_L$ values should remain restricted to around $90^\circ$ according to the restriction (\ref{eq:cond3}) posed by the equation of cylinder free motion. However, measurements show that $\phi_L$ continues to increase. In the lower branch, $7\leqslant U^* <10.5$, $\phi_L$ remains fairly constant and close to the upper limit of the permissible range.  

 In the upper branch, the frequency of oscillation increases  while the relationship between  $\phi_L$ and $f^*$ is almost linear  as discussed earlier with regard to figure~\ref{fig:phase_lift_freq}. However, this variation of $\phi_L$ does not fully conform with the restriction posed by the equation of cylinder free motion as per (\ref{eq:cond3}), which restricts $\phi_L$ to values around $90^\circ$  in the second half of the upper branch. This discrepancy may be explained by noting that the above restriction is based on the  approximation of pure harmonic motion, which does not hold in the second half of the upper branch; as already pointed out, the latter range is characterised by considerable modulations in both the displacement and the fluid forcing.  We interpret this as the result of competing factors posed by the hydrodynamics and the equation of cylinder free motion, which leads to deviations from the ideal harmonic motion. Moreover, the variation of the mean drag phase with reduced velocity in figure~\ref{fig:phase_indirect} shows that the $\phi_D$ remains approximately constant at a value of approximately $50^\circ$, which corresponds to the $\phi_D$  value at $U^*=6.2$ where  $\phi_L$  has a value equal to approximately $90^\circ$. This is consistent with the scenario of `irregular' phase dynamics  in the second half of the upper branch discussed above.

\subsection{Upper$\leftrightarrow$lower transition -- a mode competition}
\subsubsection{Phase dynamics in the mode competition region}
\begin{figure}[t]
\begin{center}
\includegraphics[width=0.56\textwidth]{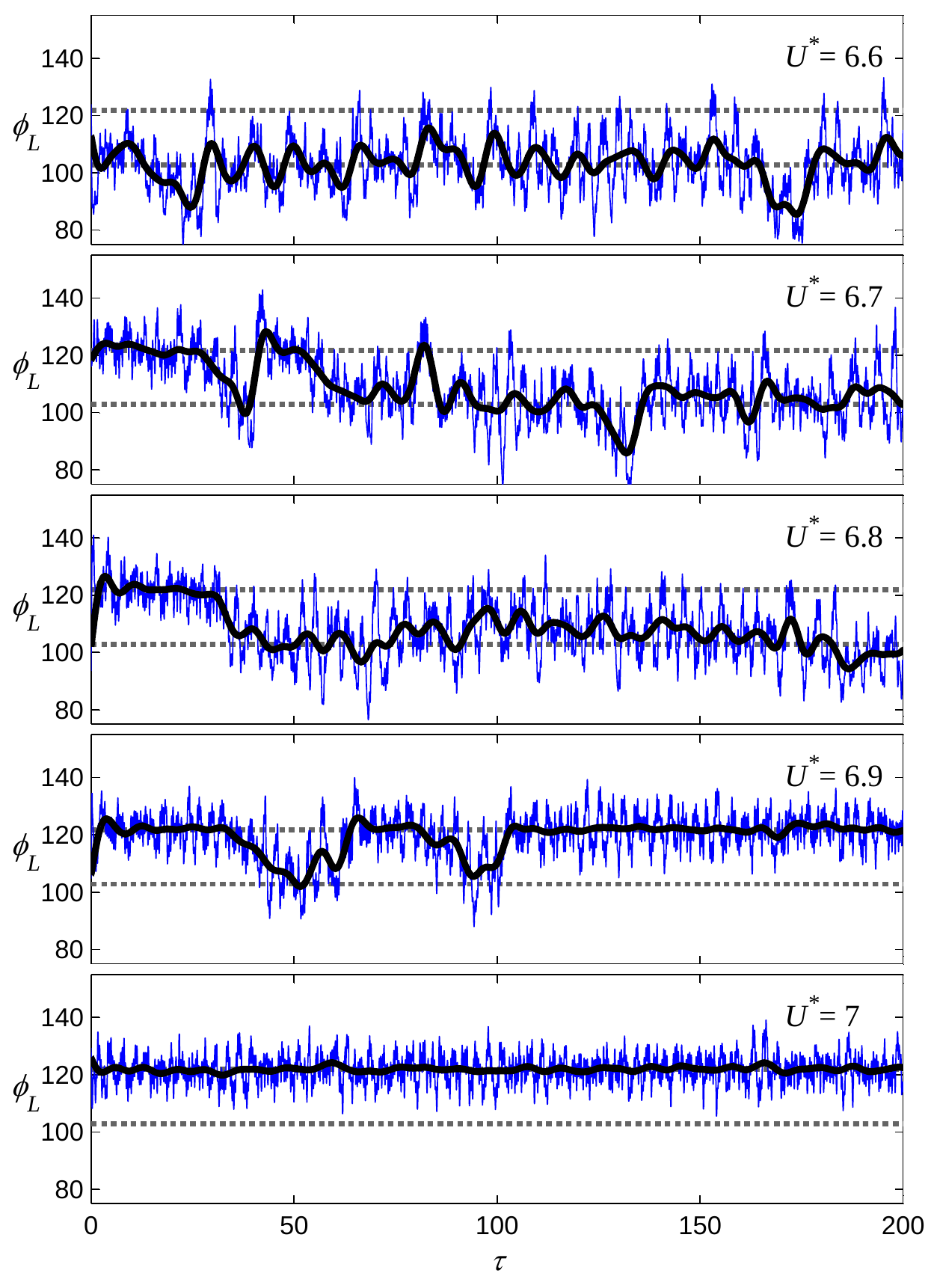}
\end{center}
\caption{(Colour online) Time series of the instantaneous phase between lift and displacement at different reduced velocities encompassing the bistable region. The blue lines denote the instantaneous phase and thick black lines denote the moving average (filtered phase).  Values are given in degrees. There is no difference between the  wrapped and the unwrapped phases for the cases shown.}\label{fig:phase_lift_bistable}
\end{figure}

The previous section has shown the appearance of chaos in the second half of the upper branch which appears to be due to the instability of a periodic limit cycle. However, we have also observed a transition region between the upper and lower branches over the range $6.7<U^*<7.0$, which we denote as the `mode competition' region; it spawns some unique characteristics that are discussed in more detail  in this section.  Figure~\ref{fig:phase_lift_bistable} presents time series of the instantaneous phase difference between lift and displacement $\phi_L$ for a sequence of five reduced velocities encompassing the mode competition region. The key here is to distinguish the character of fluctuations of the instantaneous  $\phi_L$  in the mode competition region from corresponding fluctuations that occur in the neighbouring upper and lower branches. This is facilitated by drawing horizontal lines on the plots marking two different levels at $103^\circ$ and $122^\circ$. At $U^*=6.6$, i.e.\ at the end of the upper branch just before the mode competition region, $\phi_L$ displays considerable modulations around the marked lower level. It should be remembered that at this reduced velocity the system is close to the point of  unstable equilibrium as discussed in the previous section. Although the instantaneous phase can reach the upper level, the moving average remains considerably below the  upper  level. In contrast, for reduced velocities in the bistable region, i.e.\ at $U^*=$ 6.7, 6.8, and 6.9, the moving average phase spends  portions of the  time on the upper level and some other portions at the lower level. The portion of time spend on the upper level is clearly higher at $U^*=6.9$ than at $U^*=6.7$ and 6.8. Once $U^*$ increases to 7.0, which corresponds to the start of lower branch just above the mode competition region,  $\phi_L$ fluctuates around the upper level only. Thus, the results in figure~\ref{fig:phase_lift_bistable}  indicate that for a narrow range of reduced velocities  two attractors or states co-exist. Each state corresponds to a different value of the phase difference. 

Previous studies have shown that such dynamics can result from competition between different modes of vortex shedding in the wake of a cylinder oscillating  transversely to a free stream \citep{Morse09a,Morse09b,Zhao2014a}.  More specifically, measurements of the vorticity distribution in the wake of cylinders undergoing  forced vibration  have revealed a region in the map of normalised amplitude and frequency $\{A^*:f^*/U^*\}$ where either the 2S and 2Po, or the 2Po and 2P modes of vortex shedding co-exist \citep{Morse09a}. Similar mode competition and switching has also been found in free vibration  \citep{Zhao2014a}. Therefore, the phase dynamics illustrated in this study is most probably affected by intermittent switching between different modes of vortex shedding. In particular, switching between the 2Po and 2P modes is very likely in the mode competition region.  The chaotic dynamics driven by this intermittent switching between states is distinct from the chaotic oscillations observed in the second half of the upper branch because each state also has different response characteristics as shown in table~\ref{table:bistable}. Interestingly, the magnitude of the drag and lift fluctuations is approximately the same in both states. 

\begin{table}
	\centering
	\small\addtolength{\tabcolsep}{15pt}
\begin{tabular}{lcccccc}
\hline
State & $f^*$  & $A^*$  & $C_D$ & $C_L$ & $\phi_D$ &$\phi_L$ 	\\ \hline
upper branch & 0.843 & 0.658 & 0.180 & 1.25 & $45^\circ$ & $109^\circ$\\
lower branch & 0.893 & 0.544 & 0.184 & 1.24 & $46^\circ$ & $122^\circ$\\ \hline
\end{tabular}  
		\caption{Properties of the cylinder response and the hydrodynamic force of each state in the bistable region at $U^*=6.8$.}\label{table:bistable}
\end{table}

The two competing states in the mode competition region can be further distinguished by looking at the response frequencies of each state, i.e.
\begin{eqnarray*}
\text{end of upper branch:} \qquad f^*=1.160 \Leftrightarrow  f=0.997f_n,\\
\text{start of lower branch:} \qquad f^*=1.249 \Leftrightarrow  f=1.073f_n.
\end{eqnarray*}
It should be noted that $f^*$ is the response frequency normalised by the natural frequency determined from free decay tests in still water ($f_{n,\text{water}}$) while we take the natural  frequency of the system ($f_n$) to be the  one in vacuum, which can be approximated by the value determined from free decay tests in still air ($f_{n,\text{air}}$), i.e. 
\begin{equation}
	\frac{f}{f_n} \approx \frac{f_{n,\text{water}}}{f_{n,\text{air}}}\frac{f}{f_{n,\text{water}}}=0.859f^*.
\end{equation}
Thus, the response frequency is lower than the natural frequency of the hydro-elastic cylinder in the upper branch whereas it is higher  in the lower branch.

\subsubsection{The phase jump at upper$\leftrightarrow$lower transition}
The upper$\leftrightarrow$lower transition involves a jump in both the response amplitude and frequency as well as in the phase difference between lift and displacement. The origin of these jumps can be analysed with the aid of the hydrodynamic model of drag and lift components and the equation of cylinder free motion. Now, the component of the transverse force in-phase with displacement can be obtained from the equation of cylinder motion, assuming sinusoidal oscillation, which yields 
\begin{equation}\label{eq:Cycos}
	C_y\cos\phi = 2\pi^3(m^*+1)\frac{A^*}{U^{*2}}\left[1-\left(\frac{f}{f_n}\right)^2\right],
\end{equation}
where the reduced velocity $U^*$ is normalised using the natural frequency determined from free-decay tests in still water ($f_{n,\text{water}}$) in keeping with the presentation of the results.
It follows directly from equation (\ref{eq:Cycos}) that 
\begin{eqnarray}
C_y\cos\phi \geqslant 0 \quad &  \text{or} \quad  -90^\circ < \phi \leqslant90^\circ & \quad \text{if}  \quad f \leqslant f_n,\label{eq:ineq1}\\
C_y\cos\phi<0  \quad & \text{or} \quad  90^\circ < \phi \leqslant 270^\circ & \quad \text{if}  \quad  f>f_n.\label{eq:ineq2}
\end{eqnarray}
From equations (\ref{eq:FLy_approx}) and (\ref{eq:FDy_approx}) of the hydrodynamical model, the component of the transverse force in phase with the cylinder displacement  can be written as
\begin{equation}\label{eq:Cycos2}
	C_y\cos\phi =  C_L\cos\phi_L - \frac{\pi f^*A^*}{U^*} C_D\cos\phi_D.
\end{equation}
Thus, according to (\ref{eq:ineq1}) and (\ref{eq:ineq2}) once the oscillation frequency passes through the crossover point where $f = f_n$, the  term $C_y\cos\phi$ must change sign. In the second half of the upper branch, before the crossover point where $f \leqslant f_n$, the requirement $C_y\cos\phi \geqslant 0$ from (\ref{eq:ineq1}) cannot be satisfied because $\phi_L>90^\circ$ and both terms on the right-hand side of   (\ref{eq:Cycos2}) are negative. As discussed in the previous section, in the  second half of the upper branch  the hydrodynamics are incompatible with the equation of cylinder free motion. However, as long as $f >f_n\,$  $\phi_L$ can take values above $90^\circ$ without violating the restrictions posed by the equation of cylinder free motion in (\ref{eq:ineq2}). This is consistent with measured values in the lower branch where $\phi_L\approx122^\circ$.  

\begin{figure}
\begin{center}
\includegraphics[width=0.73\textwidth]{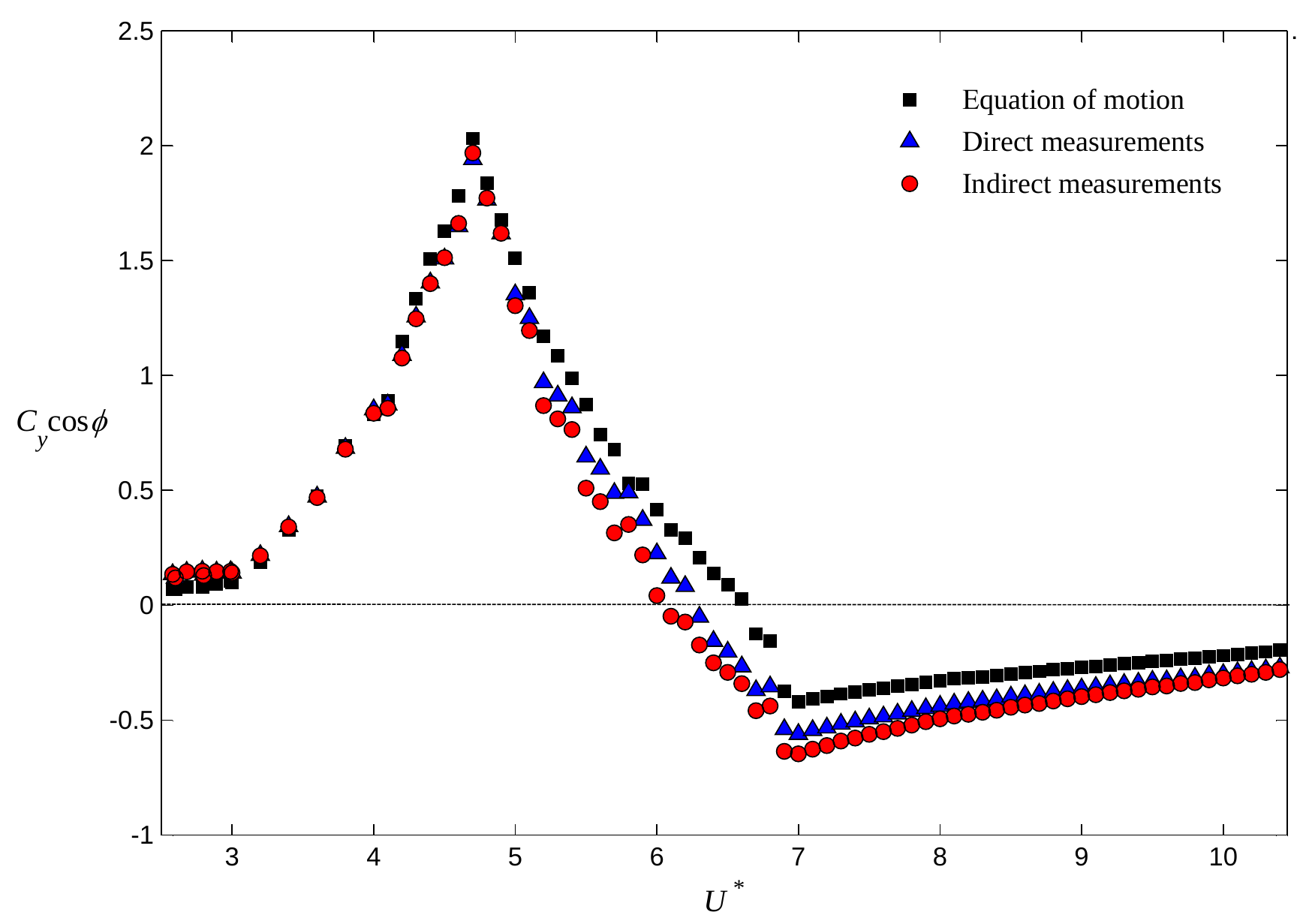}
\end{center}
\caption{(Colour online) Variation of the force component in-phase with displacement, $C_y\cos\phi$, as a function of  the reduced velocity, $U^*$, as determined by the equation of cylinder motion, direct and indirect measurements (see the text). }\label{fig:Cycos}
\end{figure}

In order to verify the above arguments, we present the variation of $C_y\cos\phi$ as a function of  $U^*$ in figure~\ref{fig:Cycos}. The first method to determine $C_y\cos\phi$ is from measurements of $A^*$ and $f^*$ and the equation of cylinder motion  (\ref{eq:Cycos}), which can be rewritten as
\begin{equation}\label{eq:Cycos3}
	C_y\cos\phi = 2\pi^3m^*\frac{A^*}{U^{*2}}\left[\left(\frac{f_{n\text{,air}}}{f_{n\text{,water}}}\right)^2-f^{*2}\right].
\end{equation}
This method assumes that the motion is pure harmonic. The second method is from direct measurements of $C_y$ and $\phi$, which are presented in \S\ref{sec:direct}. The third method is using equation (\ref{eq:Cycos2}) and indirect measurements of $C_L$,  $C_D$, $\phi_L$, and $\phi_D$,  which  are presented in \S\ref{sec:indirect}. Generally, the methods based on direct and indirect measurements agree satisfactorily over the entire range of reduced velocities with some minor deviations. This provides a self-consistency check that the data processing methods employed for obtaining the drag and lift components do not introduce considerable inaccuracies.  Both direct and indirect measurements agree well with values obtained from the equation of cylinder motion in the initial branch and in the first half of the upper branch. As expected from the equation of cylinder motion $C_y\cos\phi$ first becomes negative at $U^*\geqslant6.7$, which corresponds to the point where $f$ (time-averaged value) exceeds the natural frequency of the system, $f>f_n$. In contrast, both direct and indirect measurements show that $C_y\cos\phi$ becomes negative earlier in the middle of the upper branch $(U^*\geqslant6.0)$. It may be further noted that the difference between results from force measurements and from the equation of cylinder motion is most marked around the crossover point ($U^*=6.6$) but considerable deviations can also be observed in the lower branch $(U^*\geqslant7)$. Such deviations may be attributable to deviations of the cylinder response  from the ideal harmonic motion. 

According to the above analysis, the jump in $\phi_L$ in tandem with simultaneous jumps in $A^*$ and $f^*$  can be linked  to the elimination of the restriction  $\phi_L<90^\circ$ posed by the equation of cylinder motion for vibration frequencies $f \leqslant f_n$. Once the vibration frequency exceeds $f > f_n$, $\phi_L$ jumps at the preferred value that simultaneously satisfies both the hydrodynamics of the unsteady wake and the dynamics of  the cylinder free motion.

\section{Conclusions}
In this study we have employed a novel reconfiguration of the forces on a cylinder undergoing VIV into effective drag and lift components. In this reconfiguration, the irregular phase dynamics observed using the traditional streamwise and transverse forces are avoided, indicating the unsteady wake flow and the cylinder motion remain synchronised for all large-amplitude oscillations. We note this does not imply that the oscillations remain periodic.
We maintain that this relationship becomes clear because lift fluctuations are intrinsically linked with the vortex dynamics in the wake of the vibrating cylinder.  The instantaneous $\phi_L$ displays comparatively significant modulations in the second half of the upper branch, which is probably the cause of phenomenologically chaotic phase dynamics of the transverse force. 

An important finding of the present study is that the mean phase lag between  lift and cylinder displacement varies linearly with the vibration frequency, which is consistent with the dynamics of physical systems where frequency and phase are intrinsically related. The results also show that the drag induces a component which always opposes the cylinder velocity, i.e.\ it acts as a source of damping, whereas the lift component induces a component which drives the cylinder motion, in agreement with purely theoretical considerations.  A key finding is that the magnitude of the unsteady lift $C_L$  varies linearly with the kinematical parameter $f^*A^*/U^{*2}$, which is the  scaling of the traditional excitation coefficient in-phase with cylinder velocity $C_y\sin\phi$  expected from the equation of cylinder free motion. 

An analytical model is introduced for the unsteady lift and the steady and unsteady drag components, which explains well the observed dynamics. In the middle of the upper branch, the hydro-elastic cylinder reaches a point of unstable equilibrium where the system is most sensitive to  small perturbations. When increasing the reduced velocity in the second half of the upper branch,  there exist competing requirements resulting from (\textit{i}) the hydrodynamics of the unsteady wake, which dictates that the lift phase with respect to displacement has to increase as the oscillation frequency increases, and (\textit{ii}) the dynamics of cylinder free motion, which, assuming that the motion is pure harmonic, limits the permissible range of the lift phase to remain fixed at approximately  $90^\circ$. The competing requirements cannot both be simultaneously satisfied, suggesting the chaos in the second half of the upper branch is driven by a loss of stability of a periodic limit cycle.

For a narrow range of reduced velocity over the transition between the upper and lower branches, mode competition emerges where two states exist over different portions of time. Each state is characterised by different response amplitude, frequency, and lift phase. The two states are most probably associated with the 2Po and 2P modes of vortex formation found in the wake of oscillating cylinders \citep{Morse09a,Zhao2014a}. The chaotic dynamics in the mode competition region has distinct features from the region of chaotic vibrations at the second half of the upper branch. 

As a corollary, the force decomposition adopted in this study provides a theoretical framework for better understanding the dynamics of VIV of a rigid circular cylinder with a single degree of freedom to oscillate in the transverse direction. It would be of interest to extend this force decomposition method in the future to analyse the dynamics of the less-well understood case of  a hydro-elastic cylinder with two degrees of freedom in both streamwise and transverse directions. Furthermore, the  analytical model may be incorporated in refined semi-empirical codes for predicting VIV of long flexible cylinders. 

\section*{Acknowledgements}
The authors would like to acknowledge the financial support through Australian Research Council Discovery grants DP150103177 (JL, DL, \& JS), DP150102879 (JZ, DL, \& JS) and  DP170100275 (JZ).  



\bibliographystyle{elsarticle-harv} 
\bibliography{YJFLS_2019_730}

\begin{thebibliography}{43}
\expandafter\ifx\csname natexlab\endcsname\relax\def\natexlab#1{#1}\fi
\providecommand{\url}[1]{\texttt{#1}}
\providecommand{\href}[2]{#2}
\providecommand{\path}[1]{#1}
\providecommand{\DOIprefix}{doi:}
\providecommand{\ArXivprefix}{arXiv:}
\providecommand{\URLprefix}{URL: }
\providecommand{\Pubmedprefix}{pmid:}
\providecommand{\doi}[1]{\href{http://dx.doi.org/#1}{\path{#1}}}
\providecommand{\Pubmed}[1]{\href{pmid:#1}{\path{#1}}}
\providecommand{\bibinfo}[2]{#2}
\ifx\xfnm\relax \def\xfnm[#1]{\unskip,\space#1}\fi
\bibitem[{Bearman(1984)}]{Bearman1984}
\bibinfo{author}{Bearman, P.W.}, \bibinfo{year}{1984}.
\newblock \bibinfo{title}{Vortex shedding from oscillating bluff bodies}.
\newblock \bibinfo{journal}{Annual Review of Fluid Mechanics}
  \bibinfo{volume}{16}, \bibinfo{pages}{195--222}.
\newblock \URLprefix \url{https://doi.org/10.1146/annurev.fl.16.010184.001211},
  \DOIprefix\doi{10.1146/annurev.fl.16.010184.001211}.
\bibitem[{Bearman(2009)}]{Bearman09}
\bibinfo{author}{Bearman, P.W.}, \bibinfo{year}{2009}.
\newblock \bibinfo{title}{Understanding and predicting vortex-induced
  vibrations}.
\newblock \bibinfo{journal}{Journal of Fluid Mechanics} \bibinfo{volume}{634},
  \bibinfo{pages}{1--4}.
\bibitem[{Bearman(2011)}]{Bearman2011}
\bibinfo{author}{Bearman, P.W.}, \bibinfo{year}{2011}.
\newblock \bibinfo{title}{Circular cylinder wakes and vortex-induced
  vibrations}.
\newblock \bibinfo{journal}{Journal of Fluids and Structures}
  \bibinfo{volume}{27}, \bibinfo{pages}{648 -- 658}.
\newblock \URLprefix
  \url{http://www.sciencedirect.com/science/article/pii/S0889974611000600},
  \DOIprefix\doi{https://doi.org/10.1016/j.jfluidstructs.2011.03.021}.
\bibitem[{Blevins(2001)}]{Blevins01}
\bibinfo{author}{Blevins, R.D.}, \bibinfo{year}{2001}.
\newblock \bibinfo{title}{Flow-induced vibration}.
\newblock \bibinfo{publisher}{Krieger Publishing Company},
  \bibinfo{address}{Florida}.
\bibitem[{Blevins(2009)}]{Blevins09b}
\bibinfo{author}{Blevins, R.D.}, \bibinfo{year}{2009}.
\newblock \bibinfo{title}{Models for vortex-induced vibration of cylinders
  based on measured forces}.
\newblock \bibinfo{journal}{Journal of Fluids Engineering, Transactions ASME}
  \bibinfo{volume}{131}, \bibinfo{pages}{1--9}.
\newblock \bibinfo{note}{101203}.
\bibitem[{Blevins and Coughran(2009)}]{Blevins09a}
\bibinfo{author}{Blevins, R.D.}, \bibinfo{author}{Coughran, C.S.},
  \bibinfo{year}{2009}.
\newblock \bibinfo{title}{Experimental investigation of vortex-induced
  vibration in one and two dimensions with variable mass, damping, and
  {Reynolds} number}.
\newblock \bibinfo{journal}{Journal of Fluids Engineering, Transactions ASME}
  \bibinfo{volume}{131}, \bibinfo{pages}{1--7}.
\newblock \bibinfo{note}{101202}.
\bibitem[{{Brankovi\'c} and Bearman(2006)}]{Brankovich06}
\bibinfo{author}{{Brankovi\'c}, M.}, \bibinfo{author}{Bearman, P.W.},
  \bibinfo{year}{2006}.
\newblock \bibinfo{title}{Measurements of transverse forces on circular
  cylinders undergoing vortex-induced vibration}.
\newblock \bibinfo{journal}{Journal of Fluids and Structures}
  \bibinfo{volume}{22}, \bibinfo{pages}{829--836}.
\bibitem[{Carberry et~al.(2005)Carberry, Sheridan and Rockwell}]{Carberry2005}
\bibinfo{author}{Carberry, J.}, \bibinfo{author}{Sheridan, J.},
  \bibinfo{author}{Rockwell, D.}, \bibinfo{year}{2005}.
\newblock \bibinfo{title}{Controlled oscillations of a cylinder: forces and
  wake modes}.
\newblock \bibinfo{journal}{Journal of Fluid Mechanics} \bibinfo{volume}{538},
  \bibinfo{pages}{31--69}.
\bibitem[{Cohen(1995)}]{Cohen1995}
\bibinfo{author}{Cohen, L.}, \bibinfo{year}{1995}.
\newblock \bibinfo{title}{Time-frequency analysis}.
\newblock \bibinfo{publisher}{Prentice Hall PTR}, \bibinfo{address}{New
  Jersey}.
\bibitem[{Feng(1968)}]{Feng1968}
\bibinfo{author}{Feng, C.C.}, \bibinfo{year}{1968}.
\newblock \bibinfo{title}{{The measurement of vortex induced effects in flow
  past stationary and oscillating circular and D-section cylinders}}.
\newblock \bibinfo{type}{Master's thesis}. The University of British Columbia.
\bibitem[{Gabbai and Benaroya(2005)}]{Gabbai2005}
\bibinfo{author}{Gabbai, R.D.}, \bibinfo{author}{Benaroya, H.},
  \bibinfo{year}{2005}.
\newblock \bibinfo{title}{An overview of modeling and experiments of
  vortex-induced vibration of circular cylinders}.
\newblock \bibinfo{journal}{Journal of Sound and Vibration}
  \bibinfo{volume}{282}, \bibinfo{pages}{575 -- 616}.
\newblock \URLprefix
  \url{http://www.sciencedirect.com/science/article/pii/S0022460X04004845},
  \DOIprefix\doi{https://doi.org/10.1016/j.jsv.2004.04.017}.
\bibitem[{Gharib(1999)}]{Gharib1999}
\bibinfo{author}{Gharib, M.Z.}, \bibinfo{year}{1999}.
\newblock \bibinfo{title}{{Vortex-induced vibration, absence of lock-in and
  fluid force deduction}}.
\newblock \bibinfo{type}{Phd thesis}. California Institute of Technology.
\bibitem[{Gopalkrishnan({1993})}]{Gopalkrishnan93}
\bibinfo{author}{Gopalkrishnan, R.}, \bibinfo{year}{{1993}}.
\newblock \bibinfo{title}{{Vortex-induced forces on oscillating bluff
  cylinders}}.
\newblock \bibinfo{type}{Phd thesis}. {Massachusetts Institute of Technology}.
\bibitem[{Govardhan and Williamson(2000)}]{Govardhan00}
\bibinfo{author}{Govardhan, R.}, \bibinfo{author}{Williamson, C.H.K.},
  \bibinfo{year}{2000}.
\newblock \bibinfo{title}{Modes of vortex formation and frequency response of a
  freely vibrating cylinder}.
\newblock \bibinfo{journal}{Journal of Fluid Mechanics} \bibinfo{volume}{420},
  \bibinfo{pages}{85--130}.
\bibitem[{Griffin(1981)}]{Griffin81b}
\bibinfo{author}{Griffin, O.M.}, \bibinfo{year}{1981}.
\newblock \bibinfo{title}{{OTEC} cold water pipe design for problems caused by
  vortex-excited oscillations}.
\newblock \bibinfo{journal}{Ocean Engineering} \bibinfo{volume}{8},
  \bibinfo{pages}{129--209}.
\bibitem[{Hover et~al.(1998)Hover, Techet and Triantafyllou}]{Hover1998}
\bibinfo{author}{Hover, F.S.}, \bibinfo{author}{Techet, A.H.},
  \bibinfo{author}{Triantafyllou, M.S.}, \bibinfo{year}{1998}.
\newblock \bibinfo{title}{Forces on oscillating uniform and tapered cylinders
  in crossflow}.
\newblock \bibinfo{journal}{Journal of Fluid Mechanics} \bibinfo{volume}{363},
  \bibinfo{pages}{97--114}.
\newblock \DOIprefix\doi{10.1017/S0022112098001074}.
\bibitem[{Khalak and Williamson(1996)}]{Khalak96}
\bibinfo{author}{Khalak, A.}, \bibinfo{author}{Williamson, C.H.K.},
  \bibinfo{year}{1996}.
\newblock \bibinfo{title}{Dynamics of a hydroelastic cylinder with very low
  mass and damping}.
\newblock \bibinfo{journal}{Journal of Fluids and Structures}
  \bibinfo{volume}{10}, \bibinfo{pages}{455--472}.
\bibitem[{Khalak and Williamson(1997)}]{Khalak97a}
\bibinfo{author}{Khalak, A.}, \bibinfo{author}{Williamson, C.H.K.},
  \bibinfo{year}{1997}.
\newblock \bibinfo{title}{Fluid forces and dynamics of a hydroelastic structure
  with very low mass and damping}.
\newblock \bibinfo{journal}{Journal of Fluids and Structures}
  \bibinfo{volume}{11}, \bibinfo{pages}{973 -- 982}.
\newblock \DOIprefix\doi{https://doi.org/10.1006/jfls.1997.0110}.
\bibitem[{Khalak and Williamson(1999)}]{Khalak99}
\bibinfo{author}{Khalak, A.}, \bibinfo{author}{Williamson, C.H.K.},
  \bibinfo{year}{1999}.
\newblock \bibinfo{title}{Motions, forces and mode transitions in
  vortex-induced vibrations at low mass-damping}.
\newblock \bibinfo{journal}{Journal of Fluids and Structures}
  \bibinfo{volume}{13}, \bibinfo{pages}{813--851}.
\bibitem[{King(1977)}]{King1977}
\bibinfo{author}{King, R.}, \bibinfo{year}{1977}.
\newblock \bibinfo{title}{A review of vortex shedding research and its
  application}.
\newblock \bibinfo{journal}{Ocean Engineering} \bibinfo{volume}{4},
  \bibinfo{pages}{141 -- 171}.
\newblock \URLprefix
  \url{http://www.sciencedirect.com/science/article/pii/0029801877900026},
  \DOIprefix\doi{https://doi.org/10.1016/0029-8018(77)90002-6}.
\bibitem[{Klamo et~al.(2006)Klamo, Leonard and Roshko}]{Klamo06}
\bibinfo{author}{Klamo, J.T.}, \bibinfo{author}{Leonard, A.},
  \bibinfo{author}{Roshko, A.}, \bibinfo{year}{2006}.
\newblock \bibinfo{title}{The effects of damping on the amplitude and frequency
  response of a freely vibrating cylinder in cross-flow}.
\newblock \bibinfo{journal}{Journal of Fluids and Structures}
  \bibinfo{volume}{22}, \bibinfo{pages}{845--856}.
\bibitem[{Konstantinidis(2013)}]{Konstantinidis2013}
\bibinfo{author}{Konstantinidis, E.}, \bibinfo{year}{2013}.
\newblock \bibinfo{title}{Apparent and effective drag for circular cylinders
  oscillating transverse to a free stream}.
\newblock \bibinfo{journal}{Journal of Fluids and Structures}
  \bibinfo{volume}{39}, \bibinfo{pages}{418 -- 426}.
\newblock \URLprefix
  \url{http://www.sciencedirect.com/science/article/pii/S0889974613000728},
  \DOIprefix\doi{https://doi.org/10.1016/j.jfluidstructs.2013.03.001}.
\bibitem[{Konstantinidis(2014)}]{Konstantinidis2014}
\bibinfo{author}{Konstantinidis, E.}, \bibinfo{year}{2014}.
\newblock \bibinfo{title}{On the response and wake modes of a cylinder
  undergoing streamwise vortex-induced vibration}.
\newblock \bibinfo{journal}{Journal of Fluids and Structures}
  \bibinfo{volume}{45}, \bibinfo{pages}{256 -- 262}.
\newblock \URLprefix
  \url{http://www.sciencedirect.com/science/article/pii/S0889974613002570},
  \DOIprefix\doi{https://doi.org/10.1016/j.jfluidstructs.2013.11.013}.
\bibitem[{Konstantinidis(2017)}]{Konstantinidis2017}
\bibinfo{author}{Konstantinidis, E.}, \bibinfo{year}{2017}.
\newblock \bibinfo{title}{Comment on “lock-in in forced vibration of a
  circular cylinder” [phys. fluids 28, 113605 (2016)]}.
\newblock \bibinfo{journal}{Physics of Fluids} \bibinfo{volume}{29},
  \bibinfo{pages}{109101}.
\newblock \URLprefix \url{https://doi.org/10.1063/1.5006495},
  \DOIprefix\doi{10.1063/1.5006495}.
\bibitem[{Konstantinidis and Bouris(2016)}]{Konstantinidis2016}
\bibinfo{author}{Konstantinidis, E.}, \bibinfo{author}{Bouris, D.},
  \bibinfo{year}{2016}.
\newblock \bibinfo{title}{Vortex synchronization in the cylinder wake due to
  harmonic and non-harmonic perturbations}.
\newblock \bibinfo{journal}{Journal of Fluid Mechanics} \bibinfo{volume}{804},
  \bibinfo{pages}{248–277}.
\newblock \DOIprefix\doi{10.1017/jfm.2016.527}.
\bibitem[{Konstantinidis and Liang(2011)}]{Konstantinidis11}
\bibinfo{author}{Konstantinidis, E.}, \bibinfo{author}{Liang, C.},
  \bibinfo{year}{2011}.
\newblock \bibinfo{title}{Dynamic response of a turbulent cylinder wake to
  sinusoidal inflow perturbations across the vortex lock-on range}.
\newblock \bibinfo{journal}{Physics of Fluids} \bibinfo{volume}{23}.
\newblock \bibinfo{note}{075102}.
\bibitem[{Lee and Bernitsas(2011)}]{Lee2011}
\bibinfo{author}{Lee, J.}, \bibinfo{author}{Bernitsas, M.},
  \bibinfo{year}{2011}.
\newblock \bibinfo{title}{High-damping, high-{Reynolds VIV} tests for energy
  harnessing using the {VIVACE} converter}.
\newblock \bibinfo{journal}{Ocean Engineering} \bibinfo{volume}{38},
  \bibinfo{pages}{1697 -- 1712}.
\newblock \URLprefix
  \url{http://www.sciencedirect.com/science/article/pii/S002980181100134X},
  \DOIprefix\doi{https://doi.org/10.1016/j.oceaneng.2011.06.007}.
\bibitem[{Leontini et~al.(2006a)Leontini, Thompson and Hourigan}]{Leontini2006}
\bibinfo{author}{Leontini, J.}, \bibinfo{author}{Thompson, M.},
  \bibinfo{author}{Hourigan, K.}, \bibinfo{year}{2006}a.
\newblock \bibinfo{title}{The beginning of branching behaviour of
  vortex-induced vibration during two-dimensional flow}.
\newblock \bibinfo{journal}{Journal of Fluids and Structures}
  \bibinfo{volume}{22}, \bibinfo{pages}{857 -- 864}.
\newblock \URLprefix
  \url{http://www.sciencedirect.com/science/article/pii/S088997460600051X},
  \DOIprefix\doi{https://doi.org/10.1016/j.jfluidstructs.2006.04.003}.
\bibitem[{Leontini et~al.(2006b)Leontini, Stewart, Thompson and
  Hourigan}]{Leontini2006b}
\bibinfo{author}{Leontini, J.S.}, \bibinfo{author}{Stewart, B.E.},
  \bibinfo{author}{Thompson, M.C.}, \bibinfo{author}{Hourigan, K.},
  \bibinfo{year}{2006}b.
\newblock \bibinfo{title}{Wake state and energy transitions of an oscillating
  cylinder at low {Reynolds} number}.
\newblock \bibinfo{journal}{Physics of Fluids} \bibinfo{volume}{18},
  \bibinfo{pages}{067101}.
\newblock \URLprefix \url{https://aip.scitation.org/doi/abs/10.1063/1.2204632},
  \DOIprefix\doi{10.1063/1.2204632}.
\bibitem[{Lucor et~al.(2005)Lucor, Foo and Karniadakis}]{Lucor2005}
\bibinfo{author}{Lucor, D.}, \bibinfo{author}{Foo, J.},
  \bibinfo{author}{Karniadakis, G.E.}, \bibinfo{year}{2005}.
\newblock \bibinfo{title}{Vortex mode selection of a rigid cylinder subject to
  {VIV} at low mass-damping}.
\newblock \bibinfo{journal}{Journal of Fluids and Structures}
  \bibinfo{volume}{20}, \bibinfo{pages}{483--503}.
\bibitem[{Marzouk(2011)}]{Marzouk2011}
\bibinfo{author}{Marzouk, O.A.}, \bibinfo{year}{2011}.
\newblock \bibinfo{title}{One-way and two-way couplings of {CFD} and structural
  models and application to the wake-body interaction}.
\newblock \bibinfo{journal}{Applied Mathematical Modelling}
  \bibinfo{volume}{35}, \bibinfo{pages}{1036 -- 1053}.
\newblock \URLprefix
  \url{http://www.sciencedirect.com/science/article/pii/S0307904X10002982},
  \DOIprefix\doi{https://doi.org/10.1016/j.apm.2010.07.049}.
\bibitem[{Morse and Williamson(2009a)}]{Morse09a}
\bibinfo{author}{Morse, T.L.}, \bibinfo{author}{Williamson, C.H.K.},
  \bibinfo{year}{2009}a.
\newblock \bibinfo{title}{Fluid forcing, wake modes, and transitions for a
  cylinder undergoing controlled oscillations}.
\newblock \bibinfo{journal}{Journal of Fluids and Structures}
  \bibinfo{volume}{25}, \bibinfo{pages}{697--712}.
\bibitem[{Morse and Williamson(2009b)}]{Morse09b}
\bibinfo{author}{Morse, T.L.}, \bibinfo{author}{Williamson, C.H.K.},
  \bibinfo{year}{2009}b.
\newblock \bibinfo{title}{Prediction of vortex-induced vibration response by
  employing controlled motion}.
\newblock \bibinfo{journal}{Journal of Fluid Mechanics} \bibinfo{volume}{634},
  \bibinfo{pages}{5--39}.
\bibitem[{Morse and Williamson(2010)}]{Morse10}
\bibinfo{author}{Morse, T.L.}, \bibinfo{author}{Williamson, C.H.K.},
  \bibinfo{year}{2010}.
\newblock \bibinfo{title}{Steady, unsteady and transient vortex-induced
  vibration predicted using controlled motion data}.
\newblock \bibinfo{journal}{Journal of Fluid Mechanics} \bibinfo{volume}{649},
  \bibinfo{pages}{429–451}.
\newblock \DOIprefix\doi{10.1017/S002211200999379X}.
\bibitem[{Nemes et~al.(2012)Nemes, Zhao, Lo~Jacono and Sheridan}]{Nemes2012}
\bibinfo{author}{Nemes, A.}, \bibinfo{author}{Zhao, J.},
  \bibinfo{author}{Lo~Jacono, D.}, \bibinfo{author}{Sheridan, J.},
  \bibinfo{year}{2012}.
\newblock \bibinfo{title}{The interaction between flow-induced vibration
  mechanisms of a square cylinder with varying angles of attack}.
\newblock \bibinfo{journal}{Journal of Fluid Mechanics} \bibinfo{volume}{710},
  \bibinfo{pages}{102–130}.
\newblock \DOIprefix\doi{10.1017/jfm.2012.353}.
\bibitem[{Parkinson(1971)}]{Parkinson1971}
\bibinfo{author}{Parkinson, G.V.}, \bibinfo{year}{1971}.
\newblock \bibinfo{title}{Wind-induced instability of structures}.
\newblock \bibinfo{journal}{Philosophical Transactions of the Royal Society of
  London. Series A, Mathematical and Physical Sciences} \bibinfo{volume}{269},
  \bibinfo{pages}{395--413}.
\newblock \URLprefix
  \url{https://royalsocietypublishing.org/doi/abs/10.1098/rsta.1971.0040},
  \DOIprefix\doi{10.1098/rsta.1971.0040}.
\bibitem[{{Pikovsky} et~al.(2001){Pikovsky}, {Rosenblum}, {Kurths} and
  {Strogatz}}]{Pikovsky2001}
\bibinfo{author}{{Pikovsky}, A.}, \bibinfo{author}{{Rosenblum}, M.},
  \bibinfo{author}{{Kurths}, J.}, \bibinfo{author}{{Strogatz}, S.},
  \bibinfo{year}{2001}.
\newblock \bibinfo{title}{{Synchronization: A Universal Concept in Nonlinear
  Sciences}}.
\newblock \bibinfo{publisher}{Cambridge University Press}.
\bibitem[{Sarpkaya(2004)}]{Sarpkaya04}
\bibinfo{author}{Sarpkaya, T.}, \bibinfo{year}{2004}.
\newblock \bibinfo{title}{A critical review of the intrinsic nature nature of
  vortex-induced vibrations}.
\newblock \bibinfo{journal}{Journal of Fluids and Structures}
  \bibinfo{volume}{19}, \bibinfo{pages}{389--447}.
\bibitem[{Soti et~al.(2018)Soti, Zhao, Thompson, Sheridan and
  Bhardwaj}]{Soti18}
\bibinfo{author}{Soti, A.K.}, \bibinfo{author}{Zhao, J.},
  \bibinfo{author}{Thompson, M.C.}, \bibinfo{author}{Sheridan, J.},
  \bibinfo{author}{Bhardwaj, R.}, \bibinfo{year}{2018}.
\newblock \bibinfo{title}{Damping effects on vortex-induced vibration of a
  circular cylinder and implications for power extraction}.
\newblock \bibinfo{journal}{Journal of Fluids and Structures}
  \bibinfo{volume}{81}, \bibinfo{pages}{289 -- 308}.
\newblock \URLprefix
  \url{http://www.sciencedirect.com/science/article/pii/S0889974617305595},
  \DOIprefix\doi{https://doi.org/10.1016/j.jfluidstructs.2018.04.013}.
\bibitem[{Tanida et~al.(1973)Tanida, Okajima and Watanabe}]{Tanida1973}
\bibinfo{author}{Tanida, Y.}, \bibinfo{author}{Okajima, A.},
  \bibinfo{author}{Watanabe, Y.}, \bibinfo{year}{1973}.
\newblock \bibinfo{title}{Stability of a circular cylinder oscillating in
  uniform flow or in a wake}.
\newblock \bibinfo{journal}{Journal of Fluid Mechanics} \bibinfo{volume}{61},
  \bibinfo{pages}{769–784}.
\newblock \DOIprefix\doi{10.1017/S0022112073000935}.
\bibitem[{Williamson and Govardhan(2004)}]{Williamson04}
\bibinfo{author}{Williamson, C.H.K.}, \bibinfo{author}{Govardhan, R.},
  \bibinfo{year}{2004}.
\newblock \bibinfo{title}{Vortex-induced vibrations}.
\newblock \bibinfo{journal}{Annual Review of Fluid Mechanics}
  \bibinfo{volume}{36}, \bibinfo{pages}{413--455}.
\bibitem[{Zhao et~al.(2014a)Zhao, Leontini, Lo~Jacono and Sheridan}]{Zhao2014a}
\bibinfo{author}{Zhao, J.}, \bibinfo{author}{Leontini, J.},
  \bibinfo{author}{Lo~Jacono, D.}, \bibinfo{author}{Sheridan, J.},
  \bibinfo{year}{2014}a.
\newblock \bibinfo{title}{Chaotic vortex induced vibrations}.
\newblock \bibinfo{journal}{Physics of Fluids} \bibinfo{volume}{26},
  \bibinfo{pages}{121702}.
\newblock \URLprefix \url{https://doi.org/10.1063/1.4904975},
  \DOIprefix\doi{10.1063/1.4904975}.
\bibitem[{Zhao et~al.(2014b)Zhao, Leontini, Lo~Jacono and Sheridan}]{Zhao2014b}
\bibinfo{author}{Zhao, J.}, \bibinfo{author}{Leontini, J.},
  \bibinfo{author}{Lo~Jacono, D.}, \bibinfo{author}{Sheridan, J.},
  \bibinfo{year}{2014}b.
\newblock \bibinfo{title}{Fluid–structure interaction of a square cylinder at
  different angles of attack}.
\newblock \bibinfo{journal}{Journal of Fluid Mechanics} \bibinfo{volume}{747},
  \bibinfo{pages}{688–721}.
\newblock \DOIprefix\doi{10.1017/jfm.2014.167}.

\end{thebibliography}






\end{document}